\documentclass[aps,prl,twocolumn,showpacs,superscriptaddress, preprintnumbers]{revtex4-1}
\usepackage{latexsym}
\usepackage{amssymb}
\usepackage{graphicx}
\usepackage{amsmath}
\usepackage{bm}
\usepackage[colorlinks,
          linkcolor=black,
            citecolor=black,
            urlcolor=blue
           ]{hyperref}
\usepackage{verbatim}
\usepackage{mathrsfs}
\usepackage{extarrows}
\usepackage{comment}
\usepackage{mathtools,slashed}
\usepackage{soul}
\usepackage[toc,page]{appendix}
\usepackage[vcentermath]{youngtab}
\usepackage{multirow}
\usepackage{atbegshi,picture}
\usepackage{lipsum}
\usepackage{bbm}

\newcommand{\calH}{\mathcal{H}}
\newcommand{\tilcalH}{\tilde{\mathcal{H}}}

\DeclareMathOperator{\Tr}{Tr}

\begin{document}

\title{{
  Twisted boundary condition and Lieb-Schultz-Mattis ingappability \\ for discrete symmetries}}

\author{Yuan Yao}
\affiliation{Institute for Solid State Physics, The University of Tokyo. Kashiwa, Chiba 277-8581, Japan}

\author{Masaki Oshikawa}
\affiliation{Institute for Solid State Physics, The University of Tokyo. Kashiwa, Chiba 277-8581, Japan}
\affiliation{Kavli Institute for the Physics and Mathematics of the Universe (WPI),
The University of Tokyo, Kashiwa, Chiba 277-8583, Japan}
\affiliation{Trans-scale Quantum Science Institute, University of Tokyo, Bunkyo-ku, Tokyo 113-0033, Japan}

\date{\today}

\begin{abstract}

We discuss quantum many-body systems with lattice translation and discrete onsite symmetries.
We point out that, under a boundary condition twisted by a symmetry operation, there is an exact degeneracy of ground states if the unit cell forms a projective representation of the onsite discrete symmetry.
Based on the quantum transfer matrix formalism, we show that, if the system is gapped, the ground-state degeneracy under the twisted boundary condition also implies a ground-state (quasi-)degeneracy under the periodic boundary conditions.
This gives a compelling evidence for the recently proposed Lieb-Schultz-Mattis type ingappability due to the onsite discrete symmetry in two and higher dimensions.
\end{abstract}

\maketitle

\textit{Introduction.---}
% In condensed matter and statistical physics, identifying various quantum phases is a central,
% but in general difficult task due to complicated interactions and strong correlations.
%Symmetries often play an important role in quantifying the phase.
One of the most notable principles in quantum many-body systems is Lieb-Schultz-Mattis (LSM) theorem~\cite{Lieb:1961aa} and its generalizations~\cite{Affleck:1986aa,OYA1997,Oshikawa:2000aa,Hastings:2004ab,NachtergaeleSims}, which state an ``ingappability'' --- either the presence of gapless excitations above ground state(s) or a ground-state degeneracy in the large system size limit --- of the systems possessing U(1) and translation symmetries with a fractional filling.
Recently, {stronger constraints due to higher symmetries such as the SU(2)~\cite{Furuya:2017aa,Metlitski:2018aa} and the SU(N)~\cite{Yao:2019aa} symmetries are also investigated.}

In these variations of the LSM-type theorems, a continuous symmetry of the Hamiltonian appears essential.
However, recently, generalizations of the LSM theorem have been made for systems with only discrete symmetries~\cite{Chen-Gu-Wen_classification2010,Fuji-SymmetryProtection-PRB2016,Watanabe:2015aa,Ogata:2018aa,Ogata:2020aa}.
The discrete-symmetry version of the LSM theorem was demonstrated first based on Matrix Product State (MPS) formulation~\cite{Chen-Gu-Wen_classification2010}.
Since gapped ground states in one dimension can be generally described by a MPS, it was quite convincing.
Furthermore, the theorem was recently proved in a mathematically rigorous manner~~\cite{Ogata:2018aa,Ogata:2020aa} in one dimension.

Generalizations to higher dimensions were proposed in Ref.~\cite{Watanabe:2015aa}. It was further extended in terms of lattice homotopy description~\cite{Po:2017aa,Else:2020aa}.
Moreover, in terms of anomaly of quantum field theory,
ingappabilities related to discrete symmetries in two-dimensional systems are formulated in~\cite{Cheng:2016aa}.
However, in two and higher dimensions, the statements are not as convincingly established as in one dimension~\footnote{
The argument based on the Schmidt decomposition in Ref.~\cite{Watanabe:2015aa} is essentially one-dimensional. Its validity in higher dimensions may be questionable, especially
since there are symmetry-protected topological (SPT) phases which have a unique gapped ground state with gapless entanglement spectrum,
such as the $S=2$ Affleck-Kennedy-Lieb-Tasaki state~\cite{CPSV-EntSpec_PRB2011}.
The other argument in Ref.~\cite{Watanabe:2015aa} is based on the observation that (in the simplest case) the system containing an odd number of spins has exactly degenerate ground states.
While it is interesting, by itself it would not be sufficient to establish the statement, as the degeneracy is rather trivial.
Both arguments also depends on the special choice of the system size, which we can avoid by using the tilted boundary condition.
The arguments in later works~\cite{Cheng:2016aa,Po:2017aa,Else:2020aa} either depend on Ref.~\cite{Watanabe:2015aa}, or are abstract ones not directly on concrete lattice models.}.
Therefore, in this Letter, we provide a simple argument which strongly implies the LSM-type ingappability for discrete symmetries in higher dimensions.
The key notion is the robustness of the system to a twisting of the boundary condition by a (discrete) symmetry operation.

% This is so far limited to one dimension where the ground state of a generic quantum many-body Hamiltonian can be described by  or satisfy a separability condition
% \mycomment{Although the higher-dimensional generalizations are conjectured by a lattice homotopy description~\cite{Po:2017aa} generalized by~\cite{Else:2020aa}, 
% the related quantitative arguments largely depend on artificial system-size choices~\cite{Watanabe:2015aa}.}
% However, an explicit lattice approach is still lacking.
% Thus, it is of both practical and conceptual importances to study universal LSM-type ingappability constraints imposed by discrete symmetries in arbitrary dimensions
% In this Letter, we discuss a generalization of the LSM ingappability for quantum many-body systems with discrete symmetries, in arbitrary dimension.

\textit{Translation invariant SU(2) spin chains with a $\mathbb{Z}_2~\times~\mathbb{Z}_2$ symmetry under a symmetry-twisted boundary condition.---}
To illustrate our point, let us first consider (1+1)d SU(2) spin chains as an example. The Hamiltonian is assumed to have the global (onsite) $\mathbb{Z}_2\times\mathbb{Z}_2$ and the lattice translation symmetries.
Considering a finite chain of length $L$ with the periodic boundary condition (PBC), the symmetry generators are given by $R^\pi_{x,z}$ and $T$, where
\begin{eqnarray}
\label{dihedral}
&&R_x^\pi=\exp\left(i\pi\sum_{k=1}^LS_k^x\right);\,\,\,R_z^\pi=\exp\left(i\pi\sum_{k=1}^LS_k^z\right), \\
\label{translation}
&&{T}^{-1}\vec{S}_i{T}=\vec{S}_{i+1},\text{ if $1\leq i<L$};\,\,\,{{T}^{-1}\vec{S}_L{T}=\vec{S}_{1}},
\end{eqnarray}
where $\vec{S}_{i}\equiv[S^x_i,S^y_i,S^z_i]$ is the spin-$s$ operator at the $i$-th {site}.
A typical example of such a system is the $XYZ$ chain:
\begin{align}\label{xyz}
\calH_{XYZ} =&
  \sum_{i}J_X{S}^x_i{S}^x_{i+1}+J_Y{S}^y_i{S}^y_{i+1}+J_Z{S}^z_i{S}^z_{i+1} .
\end{align}
Instead of PBC,
now we introduce the following boundary condition, twisted by the symmetry generator $R^\pi_z$:
\begin{eqnarray}
\label{tbc}
\vec{S}_{L+k}\equiv\left(R_z^\pi\right)^{-1}\vec{S}_kR_z^\pi,\,\,\,k=1,2,\cdots,L,
\end{eqnarray}
which is periodic up to a $\pi$-rotation around the $z$-axis.
This symmetry-twisted boundary condition (STBC) can be implemented by introducing a ``twisted'' bond in any location, say between sites $L$ and $1$:
\begin{equation}
\begin{aligned}
  \calH_{XYZ}^{\text{STBC}} =&
    \sum_{i=1}^{L-1}
    \left(
    J_X{S}^x_i{S}^x_{i+1}+J_Y{S}^y_i{S}^y_{i+1}+J_Z{S}^z_i{S}^z_{i+1} 
    \right)
    \\
    &
    - J_X{S}^x_L {S}^x_1 - J_Y{S}^y_L {S}^y_1
    + J_Z{S}^z_L {S}^z_1 .
\end{aligned}
\end{equation}
This Hamiltonian still posesses the global $\mathbb{Z}_2\times\mathbb{Z}_2$ with respect to the same symmetry generators $R^\pi_{x,z}$. On the other hand, the lattice translation must be followed by a ``gauge transformation'' to leave the Hamiltonian invariant. Namely, the symmetry generator of the lattice translation should be modified to
\begin{align}\label{1d_comm}
  \tilde{T} =& e^{ i \pi S^z_1}T .
\end{align}
Now, among the three symmetry generators of the {STBC} Hamiltonian, the two satisfy the nontrivial commutation relation
\begin{align}\label{commu}
  R^\pi_x \tilde{T} =& (-1)^{2s} \tilde{T} R^\pi_x,
\end{align}
which was also obtained in the discussion of spectrum properties of Heisenberg spin chains~\cite{Hirano:2008aa}.
The phase factor $(-1)^{2s}$, which essentially comes from
\begin{align}\label{comm_su2}
  e^{i\pi S^x_1}e^{i\pi S^z_1} = & (-1)^{2s}e^{i\pi S^z_1}e^{i\pi S^x_1},
\end{align}
is nontrivial ``$-1$'' for an half-odd-integer spin $s$, while it is trivial ``$1$'' for an integer spin $s$.
{As a direct consequence of the nontrivial commutation relation~(\ref{commu}), for a half-odd-integer spin $s$, all the eigenstates of the {STBC} Hamiltonian are doubly degenerate.
It is because, given any energy eigenstate $|\Phi\rangle$ of the STBC Hamiltonian in the $\tilde{T}$-diagonalized basis, $R^\pi_x|\Phi\rangle$ must be another distinct state with the same energy but, due to Eq.~(\ref{commu}), a different $\tilde{T}$ eigenvalue if $(-1)^{2s}\neq1$, i.e., half-integer spins.} 
In particular, the ground states are doubly degenerate. This is an exact and rigorous statement.

\textit{Generalization to higher dimensions and physical consequences.---}
This analysis can be extended to translation-invariant and $\mathbb{Z}_2\times\mathbb{Z}_2$ symmetric quantum spin systems in higher dimensions. If we consider the system on a hypercubic lattice of the size $L_1 \times L_2 \times \ldots \times L_d$, we can twist the boundary condition along $1$ direction by the symmetry generator $R^\pi_z$. Such a system is invariant under the modified lattice translation operator
\begin{align}
\label{su2twist}
  \tilde{T}_1 =& T_1 \exp{\left( i \pi \sum_{\protect\vec{r}| r_1=1} S^z_{\vec{r}} \right)   } ,
\end{align}
where the summation is taken over the sites $\vec{r}$ which has the first component $r_1=1$.
{Naively imposing periodic boundary conditions along the remaining $2$ to $d$ directions, we then find the commutation relation 
$R^\pi_x \tilde{T}_1 =(-1)^{2s A} \tilde{T}_1 R^\pi_x$,} 
where $A=L_2 \times L_3 \times \ldots \times L_d$ is the number of sites per the ``cross section'' perpendicular to the twist~\cite{Watanabe:2015aa}.
All the eigenstates, including the ground states, must be exactly doubly degenerate under the STBC, if the spin $s$ is half-odd-integer and $A$ is odd (namely, all $L_2,L_3,\ldots,L_d$ are odd).
However,
it can be removed by adopting the ``tilted boundary condition'' {where the lattice translations are also included into STBC as follows~\cite{Yao:2020PRX}. Before twisting the boundary condition by $R_z^\pi$ along the {$d$-th} direction, we tilt the lattice geometry such that completion of the {$(i<d)$}-th cycle of the torus effectively increases by {$\hat{x}_{i+1}$}, i.e.
\begin{eqnarray}
(T_{i<d})^{-L_i}\vec{S}_{\vec{r}}(T_i)^{L_i}=\vec{S}_{\vec{r}+\hat{x}_{i+1}}
\end{eqnarray}
($\hat{x}_i$ the unit vector along the $i$-th direction) as sketched in FIG.~(\ref{TLBC}).
\begin{figure}[h]
\begin{center}
\includegraphics[width=8.5cm,pagebox=cropbox,clip]{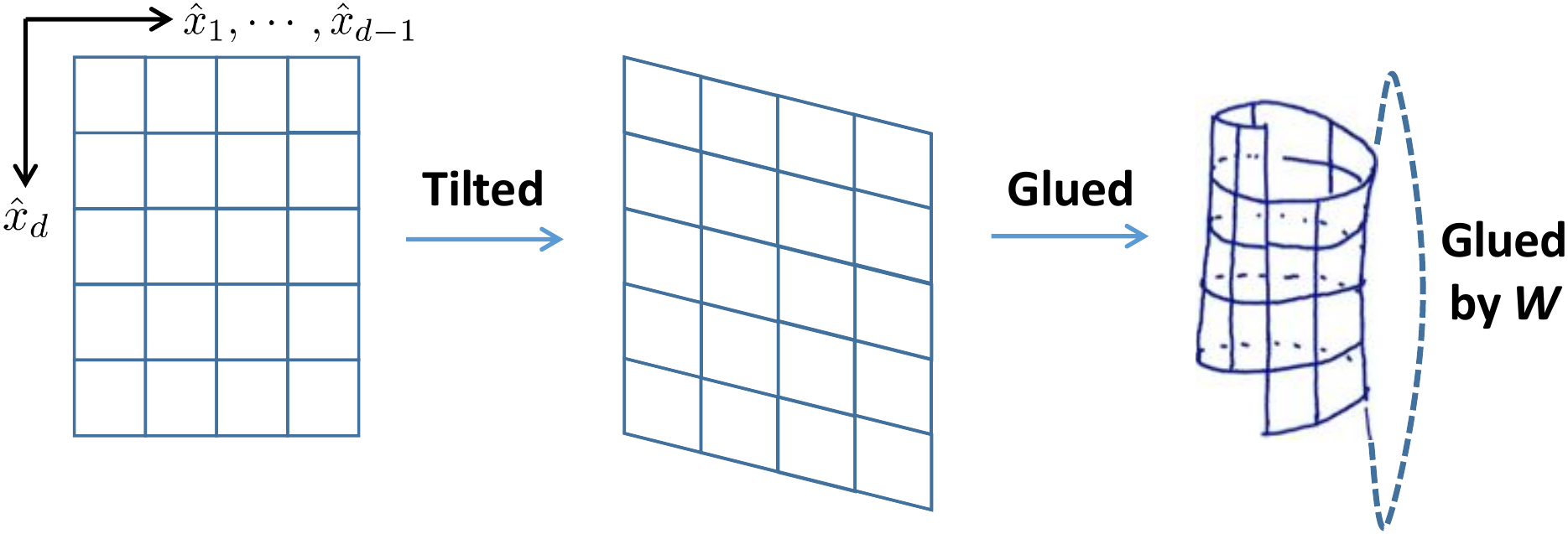}
\caption{$W$-twisted tilted boundary condition sketched for $d=2$ with $W=R_z^\pi$ for SU$(2)$ spins and $W=W_N$ for SU$(N)$. }%
\label{TLBC}
\end{center}
\end{figure}
Then we obtain a nontrivial commutation relation~\footnote{See Supplemental Materials for a diagrammatic proof and the detailed definition of $\tilde{T}_1$.} 
\begin{eqnarray}
R^\pi_x \tilde{T}_1 =(-1)^{2s} \tilde{T}_1 R^\pi_x,
\end{eqnarray}
thereby the exact double degeneracy when $s$ is half-odd-integer.}
It is also straightforward to generalize this to a more complicated systems with multiple spins within the unit cell. In this case, the exact double degeneracy holds if the total spin per unit cell is half-odd-integer, under the appropriate STBC.

Obviously, however, the above argument only applies to the system under the STBC. The question is then: is this degeneracy an unphysical artifact specific to the STBC, or is it of a physical significance?
We argue that the latter must be the case. In fact, if the system is completely disordered without any order (conventional or topological), the system must be insensitive to the symmetry-twist of the boundary condition, and thus the ground-state degeneracy should not appear under STBC. The ground-state degeneracy under the STBC then implies some kind of ingappability (gapless excitations or degenerate ground states below a non-vanishing gap).
Namely, we assert the following proposition.

Proposition:
\textit{
When a quantum many-body system with the lattice translation invariance and a global (continuous or discrete) symmetry has a gapped spectrum with a unique ground state under the PBC,
the system should also have a unique gapped ground state under the STBC.}

This proposition, together with the exact double degeneracy under the STBC implies that, a translation-invariant and $\mathbb{Z}_2\times\mathbb{Z}_2$ symmetric quantum spin system with a half-odd-integer spin per unit cell cannot have a unique gapped ground state under the PBC as well.
This amounts to be the generalization of the LSM theorem to the discrete $\mathbb{Z}_2\times\mathbb{Z}_2$. For one dimension, this was already proved within the MPS framework in Ref.~\cite{Watanabe:2015aa} and rigorously in Ref.~\cite{Ogata:2020aa}, so the present argument gives an alternative physical interpretation. On the other hand, the present result naturally leads to the same statement in arbitrary dimensions {in a firm basis.}
We note that, as in the case of the original LSM theorem~\cite{Lieb:1961aa,Affleck:1986aa,OYA1997} for the $U(1)$ symmetry, the rigorous theorem~\cite{Ogata:2020aa} can be applied to quasi one-dimensional systems for an arbtrary large, but fixed, odd cross section $A=L_2 \times L_3 \times \ldots \times L_d$.
However, here we would like to discuss general limits including $L_i \to \infty$ keeping all the $L_i$ ($i=1,\ldots,d$) to be of the same order of magnitude $O(L)$, which is the natural thermodynamic limit for the $d$-dimensional system~\cite{Oshikawa:2000aa,Hastings:2004ab,NachtergaeleSims}.

When applied to a continuous U(1) (onsite) symmetry, the STBC corresponds to insertion of an Aharonov-Bohm flux.
This also shows that the present approach is a natural generalization of the LSM theorem from the $U(1)$ to discrete symmetries.
More precisely, the above proposition applied to the case of U(1) symmetry implies that, if the gapped unique ground state is unique under the PBC, the ground state is gapped and unique for any value of Aharonov-Bohm flux inserted. This corresponds to the conjectured insensitivity of the excitation gap to the Aharonov-Bohm flux~\cite{Oshikawa:2000aa,Watanabe:2018aa} which leads to the LSM ingappability for systems with the U(1) symmetry. However, in this approach, we cannot constrain the number of the degenerate ground states (other than that it must be greater than $1$).

\textit{Robustness of the spectrum to the symmetry-twist.---}
Now let us show that the proposition naturally follows from mild assumptions.
We consider the partition function of the finite-size system of the size $L_1 \times L_2 \times \ldots L_d$ with the PBC, at the inverse temperature $\beta$. 
Let us assume that it has a unique ground state below the gap $\Delta$, {which we will show to induce contradictions}.

When the spectrum is gapped, Hastings and Koma proved that the correlation function of any local operator decays exponentially~\cite{Hastings:2006aa}. Since the system has a finite length scale (correlation length) $\xi$, we can expect the ground-state energy of the finite-size system is given in terms of the ground-state energy density $\epsilon_0$ in the thermodynamic limit, with exponentially small corrections:
\begin{align}
  E_0 =&  \epsilon_0 V +  O( e^{-L/\xi} ) ,
  \label{eq.E0_corr}
\end{align}
where $L=\min{\left(L_1,L_2,\ldots,L_d \right)}$, and $V = L_1 L_2 \ldots L_d$ is the volume of the system.
The correction might actually be of the form $O(L^\chi e^{-L/\xi})$ for some constant exponent $\chi$. However, we can cover such a case with $O(e^{-L/\xi})$ by taking a larger value of $\xi$.
Furthermore, the partition function $Z = \Tr{e^{-\beta \calH}}$ is expected to obey
\begin{align}
  \log{Z(\beta,\{L_\alpha\})} \sim  - \beta V \epsilon_0 
     +O(\beta e^{-L/\xi}) + O(V e^{-\beta \Delta}) .
\label{eq.logZ_gap}
\end{align}
The first term $O(\beta e^{-L/\xi})$ simply comes from the exponentially small correction~\eqref{eq.E0_corr} to the ground-state energy. 
The second term $O(V e^{-\beta \Delta})$ represents the contribution from excited states. Here we assume that such a contribution is exponentially small per volume.
Equation~\eqref{eq.logZ_gap} implies that the entropy in the thermodynamic limit and zero-temperature limit taken simultaneously {with $\beta \sim (L)^z$} for any ``dynamical exponent'' $z>0$ is
\begin{align}
  S_0 =& \lim_{\beta=(L)^z \to \infty}
\!\!  \left[
  \log{Z(\beta,\{L_\alpha\})}\!-\!\beta \frac{\partial}{\partial \beta}\log{Z(\beta,\{L_\alpha\})}
  \right] 
  \!\!= 0 .
\label{eq.gs_entropy}
\end{align}

Now let us consider the same system in terms of the quantum transfer matrix (QTM)~\cite{Betsuyaku-PRL1984} (For a recent application of QTM, see for example Ref.~\cite{Piroli-QTM2017}).
The partition function of the quantum system can be identified with the partition function of a classical statistical system in $d+1$ dimensions by the Lie-Trotter-Suzuki decomposition~\cite{Suzuki:1976aa}.
QTM is the transfer matrix of this classical statistical system along a spatial direction~\footnote{The mapping to the classical statistical system is exact only in the limit of an infinite Trotter number. Here we assume that the QTM is well-defined in this limit.}.
As an alternative to the conventional formulation of the QTM, we can also represent the infinitesimal imaginary time evolution operator $e^{-\delta \calH}$
(for a small but finite $\delta$ in practical calculations)
in terms of a Matrix Product Operator (MPO) which is defined by contracting constituent tensors in the spatial directions~\cite{CrossWhite-Bacon-MPO_PRA2008,Pirvu-MPO_NJP2010}.
The partition function is given by the trace of a power of the MPO, and the QTM is a different MPO obtained by contracting constituent tensors in the imaginary time direction,
leaving one spatial direction uncontracted~\footnote{See Supplemental Material for more details.}.

Let $e^{-\tilde{H}}$ be the QTM along $1$ direction, in either construction.
Using this, the partition function can be rewritten as $Z = \Tr{e^{-L_1 \tilde{H}}}$.
The asymptotic expression~\eqref{eq.logZ_gap} implies that, the spectrum of the ``QTM Hamiltonian'' $\tilcalH$ must be also gapped, with $1/\xi$ being the ``excitation gap''.
The excitation gap $\Delta$ of the original Hamiltonian $\calH$ now corresponds to the correlation length in the imaginary time direction, which plays the role of one of the spatial coordinates.
Equation~\eqref{eq.logZ_gap} and its consequence~\eqref{eq.gs_entropy} implies that the ``ground state'' $|\tilde{0}\rangle$ of the QTM Hamiltonian $\tilcalH$ is also unique. 

In the QTM formalism, the STBC is represented by an application of the symmetry operation $\tilde{U}_g$ on the ``state'' at a fixed ``time'' (the coordinate $r_1$).
The STBC corresponding to the inverse symmetry operation is represented by $\tilde{U}_g^\dagger$.
Twisting the boundary condition by a symmetry operation in one place and its inverse elsewhere can be eliminated by a gauge transformation, leaving the partition function identical to the PBC. This implies the symmetry of the QTM~\footnote{See Supplemental Material for the proofs.}
\begin{align}
  [\tilde{U}_g,\tilde{H}] =& 0 ,
\end{align}
and the unitarity of the operator
\begin{align}
  \tilde{U}_g^\dagger \tilde{U}_g = & \mathbbm{1} .
\end{align}

Now let us consider the partition function of the system under the STBC, denoted by $Z_s$. In the QTM formalism, it is given as
\begin{align}
  Z_s =& \Tr{\left( \tilde{U}_g e^{-L_1 \tilde{H}}\right) }.
\end{align}
As in the case of the PBC, the partition function in the thermodynamic and zero-temperature limit is dominated by the ``ground state'' of the QTM. The symmetry of the QTM implies that the unique ``ground state'' of the QTM is an eigenstate of the operator $\tilde{U}_g$.
Let $e^{i\theta_0}$ be the eigenvalue of $\tilde{U}_g$ of the QTM ``ground state'' $|\tilde{0}\rangle$.
Then
\begin{align}\label{unique}
  Z_s \sim e^{i\theta_0} e^{-\beta V  \epsilon_0 } ,
\end{align}
in the thermodynamic/zero-temperature limit, 
up to exponentially small corrections.
Since the Hamiltonian is Hermitian even under the STBC, the partition function $Z_s$ must be real and positive, implying $\theta_0=0$.
Thus STBC partition function $Z_s$ is essentially identical to the PBC partition function up to exponentially small corrections, implying that the system has a unique gapped ground state also for STBC since the entropy in the limit $\beta  = (L)^z \to \infty$ is zero as in the case of PBC~\eqref{eq.gs_entropy}. This is nothing but our proposition. 
{By the ``proof by contradiction''},
the ingappability of the translation-invariant and $\mathbb{Z}_2\times\mathbb{Z}_2$ symmetric spin system with a half-odd-integer spin per unit cell then follows.

{For $d=2$, the ingappability derived in the present work is consistent with
the phase diagram of the Kitaev-Heisenberg model on a honeycomb-triangular lattice, 
where featureless gapped phases are absent~\cite{Kishimoto:2018aa,Maksimov:2019aa}, 
because the number of spin-$1/2$'s per unit cell is either $1$ or $3$ for generic parameters.
On the other hand, in the Kitaev model on honeycomb
lattices~\cite{Kitaev:2006aa} where each unit cell possesses two spin-$1/2$'s, no trivial phase is found although it would be allowed by the present result. 
This is not a contradiction, however, as the inverse of the LSM-type ingappability does not always hold.
In fact, a featureless gapped ground state of spin-1/2's on a honeycomb lattice can be realized
by including farther than nearest-neighbor couplings~\cite{Kimchi-featureless-honeycomb}.}

%\bigskip
%
%\hrulefill

\textit{General LSM ingappabilities and applications.---}
We can generalize the ingappabilities above to a more general class of unitary onsite symmetry groups.
The projective nature of the onsite symmetry characterized by the commutator (\ref{comm_su2}) directly results in the ingappabilities.
$\mathbb{Z}_2\times\mathbb{Z}_2$ can be generalized to an arbitrary unitary symmetry $G$ that can be centrally extended to $G'$, e.g., $G=G'/H$ with $H$ a center subgroup of $G'$.
Then each unit cell consists of tensor products of $G'$-``spins''. 
In the SU$(2)$ case, $G'$ is arbitrary finite products of $R^\pi_x$ and $R^\pi_z$ treated as a subgroup of SU$(2)$ (rather than PSU$(2)$) and $H=\{1,(R^\pi_z)^2\}$.

Generalizing $(R_{x}^\pi,R_{z}^\pi)$ and their relation (\ref{comm_su2}) per unit cell, $G'$ can possess an almost commuting pair~\cite{Schweigert:1997aa,Borel:2002aa,Kac:2000aa} $v_{\vec{r}}$ and $w_{\vec{r}}$ satisfying:
\begin{eqnarray}\label{comm}
v_{\vec{r}}w_{\vec{r}}=h_{V,W}w_{\vec{r}}v_{\vec{r}},
\end{eqnarray}
where $h_{V,W}\in H$ is an $\vec{r}$-independent U$(1)$ phase due to the translation symmetry and Schur's Lemma~\cite{Georgi:2000}.
Parallel to the former discussions,
we can twist and tilt the boundary condition by $W=\prod_{\vec{r}}w_{\vec{r}}$ and translations as in FIG.~(\ref{TLBC}) and calculate the commutator between the modified translation
\begin{eqnarray}\label{general_transl}
\mathcal{T}_1=T_1\prod_{\vec{r}|{r_1=1}}w_{\vec{r}}, 
\end{eqnarray}
and $V=\prod_{\vec{r}}v_{\vec{r}}$ as
\begin{eqnarray}\label{ingap}
V\mathcal{T}_1=h_{V,W}\mathcal{T}_{1}V.
\end{eqnarray}
Therefore, a nontrivial $h_{V,W}=\exp(i2\pi p/q)\neq1$ (with $p$ and $q$ coprime) results in an exact $q$-fold degeneracy under STBC, which further implies an ingappability of the system under PBC, i.e. a unique gapped ground state is forbidden,
by the Proposition of the spectrum insensitivity to the symmetry-twists argued by the QTM formalism.
It can be summarized as

\textit{If a Hamiltonian on a $d$-dimensional periodic lattice possesses $G=G'/H$ symmetry and translational symmetries, a unique gapped ground state is strictly forbidden if there is a nontrivial almost commuting pair per unit cell, i.e. $h_{V,W}\neq1$ in Eq.~(\ref{comm}).}

Since such a general result and Eq.~(\ref{ingap}) are abstract, 
a further example beyond the simplest SU$(2)$ case might be helpful. 
As a natural generalization of SU$(2)$ spins,
the systems with the following SU$(N)$ degrees of freedom satisfy the su$(N)$ algebra:
\begin{eqnarray}
\label{su(n)_comm}
\left[\mathcal{S}^\beta_{\alpha,\vec{r}}\,\,,\mathcal{S}^\rho_{\sigma,\vec{r}'}\right]=\delta_{\vec{r},\vec{r}'}\left[\delta^\beta_\sigma \mathcal{S}^\rho_{\alpha,\vec{r}}-\delta^\rho_\alpha \mathcal{S}^\beta_{\sigma,\vec{r}}\right].
\end{eqnarray}
Here $\left\{\mathcal{S}^\alpha_{\beta,\vec{r}}\right\}$ defined on the unit cell at $\vec{r}$ are labelled by $1\leq\alpha\leq N$ and $1\leq\beta\leq N$,
with a constraint $\sum_{\alpha=1}^N\mathcal{S}^\alpha_\alpha=0$~\cite{Affleck:1987aa,Affleck:1988aa}.
Additionally, the full global SU(N) symmetry acts on the generators by $u\in$ SU(N):
\begin{eqnarray}
\label{su(N)_trans}
\mathcal{S}^\alpha_\beta\rightarrow(u)_{\beta,\gamma}\,\mathcal{S}^\rho_\gamma\,\,(u^\dagger)_{\rho,\alpha}
\end{eqnarray}
which does nothing if the SU(N) matrix $u$ is in the $\mathbb{Z}_N$ center of SU$(N)$: $u\propto 1$.
Thus, the physical symmetry is only PSU(N)$\cong$SU(N)/$\mathbb{Z}_N$ if the fundamental degrees of freedom consist of $\{\mathcal{S}^\alpha_\beta\}$ only.

{
Experimental realizations of the SU$(N)$ symmetry have been proposed for ultracold atoms on optical lattices~\cite{Wu:2003aa,Honerkamp:2004aa,Cazalilla:2009aa,Gorshkov:2010aa,Taie:2012aa,Pagano:2014aa,Scazza:2014aa,Zhang:2014aa}.}
Let us consider a translation-invariant SU(N)-spin Hamiltonian $\mathcal{H}$ possessing $\mathbb{Z}_N\times\mathbb{Z}_N\subset$PSU(N), which is generated by:
\begin{eqnarray}
\label{ZN_twist}
V_N&=&\prod_{\vec{r}}v_{N,\vec{r}};\,\,\,W_N=\prod_{\vec{r}}w_{N,\vec{r}},
\end{eqnarray}
of which an explicit form together with a SU$(N)$-analog of the $XYZ$ model can be found in~\footnote{See Supplemental Materials for an explicit form of $V_N$ and $W_N$, and a generalized $XYZ$ model.}. $V_N$ (also known as the shift symmetry~\cite{Shimizu:2018aa,Tanizaki:2017aa}) and $W_N$ reduce to $R^\pi_x$ and $R^\pi_z$, respectively, when $N=2$, 
but we only focus on the following essential relation
\begin{eqnarray}
\label{comm_zn}
v_{N,\vec{r}}w_{N,\vec{r}}=\exp\left(i2\pi\frac{b}{N}\right)w_{N,\vec{r}}v_{N,\vec{r}},
\end{eqnarray}
where $b$ is the total number of Young-tableau (YT) boxes~\cite{Georgi:2000} per unit cell as a generalization of $2s$ of a SU$(2)$ spin-$s$ chain.
Then, taking $v_{\vec{r}}=v_{N,\vec{r}}$ and $w_{\vec{r}}=w_{N,\vec{r}}$ in the general setting (\ref{general_transl},\ref{ingap}), 
we obtain
\begin{eqnarray}
V\mathcal{T}_1=\exp(i2\pi b/N) \mathcal{T}_1V,
\end{eqnarray}
which implies that the spin system has an exact $N/\text{gcd}(b,N)$-fold degeneracy under STBC (``gcd'': the positive greatest common divisor) and, by the insensitivity of ingappability to boundary conditions, it cannot possess a unique gapped ground state under PBC if the total YT boxes per unit is indivisible by $N$. 

\textit{Conclusions and conjectures.---}
We introduced a symmetry-twisted boundary condition in arbitrary dimensions, under which
the energy spectrum of translation invariant lattice system can be exactly degenerate.
We further argue that such an exact degeneracy implies the ingappability of ground states under PBC, based on a QTM formulation of partition functions.
While the present argument is still not mathematically rigorous, it convincingly demonstrates the LSM-type ingappability for systems
with only discrete symmetries in arbitrary dimensions.

% Our argument relies on hypotheses~\eqref{eq.E0_corr} and~\eqref{eq.logZ_gap} on the {asymptotic behaviors of the ground-state energy and free energy} of a gapped system in the low-temperature and thermodynamic limit. While we believe them to be reasonable for systems with short-range interactions only, they are not rigorously proved. We hope that the present work will stimulate further investigations in rigorous mathematics.

The present QTM-type argument does not make a full use of the {the number of degenerate ground states} under
STBC because only the case of a unique gapped ground state is excluded based on Eq.~(\ref{unique}). 
Nevertheless, it would still be reasonable to conjecture that {the same number of the degenerate ground states} is also held by the gapped ground states under PBC.
Its resolution is left as a problem for future.
We also hope that the present work will stimulate further studies on the subject, including the treatment of symmetries we did not discuss in this Letter, such as anti-unitary (time reversal) symmetries.

% Furthermore, since our treatment of symmetry twisting in addition to the QTM argument (\ref{unique}) strongly depends on the unitarity of the symmetry,
% it cannot be directly applied to the anti-unitary symmetries, e.g. time reversal.
% In contrast to Eq.~(\ref{ZN_twist}), there is no general way to factorize an anti-unitary transformation in a local way due to an additional complex conjugation.
% Nevertheless, in one dimension, the description of matrix product states allow the complex conjugation to be assigned locally~\cite{Chen:2015ab,Watanabe:2015aa},
% where we expect our method is still applicable.
% Higher-dimensional cases with anti-unitary symmetries are of future interest.

%Acknowledgements

The authors thank Hosho Katsura and Hal Tasaki for stimulating discussions, and Haruki Watanabe for informing us of related works.
Y.~Y. was supported by JSPS fellowship.
This work was supported in part by MEXT/JSPS KAKENHI Grant Nos. JP19J13783 (Y.~Y.), JP17H06462 (M.~O.), JP19H01808 (M.~O.), and JST CREST Grant No. JPMJCR19T2 (M.~O.).

%\bibliography{bib}

\begin{thebibliography}{53}%
\makeatletter
\providecommand \@ifxundefined [1]{%
 \@ifx{#1\undefined}
}%
\providecommand \@ifnum [1]{%
 \ifnum #1\expandafter \@firstoftwo
 \else \expandafter \@secondoftwo
 \fi
}%
\providecommand \@ifx [1]{%
 \ifx #1\expandafter \@firstoftwo
 \else \expandafter \@secondoftwo
 \fi
}%
\providecommand \natexlab [1]{#1}%
\providecommand \enquote  [1]{``#1''}%
\providecommand \bibnamefont  [1]{#1}%
\providecommand \bibfnamefont [1]{#1}%
\providecommand \citenamefont [1]{#1}%
\providecommand \href@noop [0]{\@secondoftwo}%
\providecommand \href [0]{\begingroup \@sanitize@url \@href}%
\providecommand \@href[1]{\@@startlink{#1}\@@href}%
\providecommand \@@href[1]{\endgroup#1\@@endlink}%
\providecommand \@sanitize@url [0]{\catcode `\\12\catcode `\$12\catcode
  `\&12\catcode `\#12\catcode `\^12\catcode `\_12\catcode `\%12\relax}%
\providecommand \@@startlink[1]{}%
\providecommand \@@endlink[0]{}%
\providecommand \url  [0]{\begingroup\@sanitize@url \@url }%
\providecommand \@url [1]{\endgroup\@href {#1}{\urlprefix }}%
\providecommand \urlprefix  [0]{URL }%
\providecommand \Eprint [0]{\href }%
\providecommand \doibase [0]{http://dx.doi.org/}%
\providecommand \selectlanguage [0]{\@gobble}%
\providecommand \bibinfo  [0]{\@secondoftwo}%
\providecommand \bibfield  [0]{\@secondoftwo}%
\providecommand \translation [1]{[#1]}%
\providecommand \BibitemOpen [0]{}%
\providecommand \bibitemStop [0]{}%
\providecommand \bibitemNoStop [0]{.\EOS\space}%
\providecommand \EOS [0]{\spacefactor3000\relax}%
\providecommand \BibitemShut  [1]{\csname bibitem#1\endcsname}%
\let\auto@bib@innerbib\@empty
%</preamble>
\bibitem [{\citenamefont {Lieb}\ \emph {et~al.}(1961)\citenamefont {Lieb},
  \citenamefont {Schultz},\ and\ \citenamefont {Mattis}}]{Lieb:1961aa}%
  \BibitemOpen
  \bibfield  {author} {\bibinfo {author} {\bibfnamefont {E.}~\bibnamefont
  {Lieb}}, \bibinfo {author} {\bibfnamefont {T.}~\bibnamefont {Schultz}}, \
  and\ \bibinfo {author} {\bibfnamefont {D.}~\bibnamefont {Mattis}},\
  }\href@noop {} {\bibfield  {journal} {\bibinfo  {journal} {Ann. Phys.}\
  }\textbf {\bibinfo {volume} {16}},\ \bibinfo {pages} {407} (\bibinfo {year}
  {1961})}\BibitemShut {NoStop}%
\bibitem [{\citenamefont {Affleck}\ and\ \citenamefont
  {Lieb}(1986)}]{Affleck:1986aa}%
  \BibitemOpen
  \bibfield  {author} {\bibinfo {author} {\bibfnamefont {I.}~\bibnamefont
  {Affleck}}\ and\ \bibinfo {author} {\bibfnamefont {E.~H.}\ \bibnamefont
  {Lieb}},\ }\href@noop {} {\bibfield  {journal} {\bibinfo  {journal} {Lett.
  Math. Phys.}\ }\textbf {\bibinfo {volume} {12}},\ \bibinfo {pages} {57}
  (\bibinfo {year} {1986})}\BibitemShut {NoStop}%
\bibitem [{\citenamefont {Oshikawa}\ \emph {et~al.}(1997)\citenamefont
  {Oshikawa}, \citenamefont {Yamanaka},\ and\ \citenamefont
  {Affleck}}]{OYA1997}%
  \BibitemOpen
  \bibfield  {author} {\bibinfo {author} {\bibfnamefont {M.}~\bibnamefont
  {Oshikawa}}, \bibinfo {author} {\bibfnamefont {M.}~\bibnamefont {Yamanaka}},
  \ and\ \bibinfo {author} {\bibfnamefont {I.}~\bibnamefont {Affleck}},\ }\href
  {\doibase 10.1103/PhysRevLett.78.1984} {\bibfield  {journal} {\bibinfo
  {journal} {Phys. Rev. Lett.}\ }\textbf {\bibinfo {volume} {78}},\ \bibinfo
  {pages} {1984} (\bibinfo {year} {1997})}\BibitemShut {NoStop}%
\bibitem [{\citenamefont {Oshikawa}(2000)}]{Oshikawa:2000aa}%
  \BibitemOpen
  \bibfield  {author} {\bibinfo {author} {\bibfnamefont {M.}~\bibnamefont
  {Oshikawa}},\ }\href {https://link.aps.org/doi/10.1103/PhysRevLett.84.1535}
  {\bibfield  {journal} {\bibinfo  {journal} {Phys. Rev. Lett.}\ }\textbf
  {\bibinfo {volume} {84}},\ \bibinfo {pages} {1535} (\bibinfo {year}
  {2000})}\BibitemShut {NoStop}%
\bibitem [{\citenamefont {Hastings}(2004)}]{Hastings:2004ab}%
  \BibitemOpen
  \bibfield  {author} {\bibinfo {author} {\bibfnamefont {M.~B.}\ \bibnamefont
  {Hastings}},\ }\href {https://link.aps.org/doi/10.1103/PhysRevB.69.104431}
  {\bibfield  {journal} {\bibinfo  {journal} {Phys. Rev. B}\ }\textbf {\bibinfo
  {volume} {69}},\ \bibinfo {pages} {104431} (\bibinfo {year}
  {2004})}\BibitemShut {NoStop}%
\bibitem [{\citenamefont {Nachtergaele}\ and\ \citenamefont
  {Sims}(2007)}]{NachtergaeleSims}%
  \BibitemOpen
  \bibfield  {author} {\bibinfo {author} {\bibfnamefont {B.}~\bibnamefont
  {Nachtergaele}}\ and\ \bibinfo {author} {\bibfnamefont {R.}~\bibnamefont
  {Sims}},\ }\href {\doibase 10.1007/s00220-007-0342-z} {\bibfield  {journal}
  {\bibinfo  {journal} {Communications in Mathematical Physics}\ }\textbf
  {\bibinfo {volume} {276}},\ \bibinfo {pages} {437} (\bibinfo {year}
  {2007})}\BibitemShut {NoStop}%
\bibitem [{\citenamefont {Furuya}\ and\ \citenamefont
  {Oshikawa}(2017)}]{Furuya:2017aa}%
  \BibitemOpen
  \bibfield  {author} {\bibinfo {author} {\bibfnamefont {S.~C.}\ \bibnamefont
  {Furuya}}\ and\ \bibinfo {author} {\bibfnamefont {M.}~\bibnamefont
  {Oshikawa}},\ }\href
  {https://link.aps.org/doi/10.1103/PhysRevLett.118.021601} {\bibfield
  {journal} {\bibinfo  {journal} {Phys. Rev. Lett.}\ }\textbf {\bibinfo
  {volume} {118}},\ \bibinfo {pages} {021601} (\bibinfo {year}
  {2017})}\BibitemShut {NoStop}%
\bibitem [{\citenamefont {Metlitski}\ and\ \citenamefont
  {Thorngren}(2018)}]{Metlitski:2018aa}%
  \BibitemOpen
  \bibfield  {author} {\bibinfo {author} {\bibfnamefont {M.~A.}\ \bibnamefont
  {Metlitski}}\ and\ \bibinfo {author} {\bibfnamefont {R.}~\bibnamefont
  {Thorngren}},\ }\href@noop {} {\bibfield  {journal} {\bibinfo  {journal}
  {Phys. Rev. B}\ }\textbf {\bibinfo {volume} {98}},\ \bibinfo {pages} {085140}
  (\bibinfo {year} {2018})}\BibitemShut {NoStop}%
\bibitem [{\citenamefont {Yao}\ \emph {et~al.}(2019)\citenamefont {Yao},
  \citenamefont {Hsieh},\ and\ \citenamefont {Oshikawa}}]{Yao:2019aa}%
  \BibitemOpen
  \bibfield  {author} {\bibinfo {author} {\bibfnamefont {Y.}~\bibnamefont
  {Yao}}, \bibinfo {author} {\bibfnamefont {C.-T.}\ \bibnamefont {Hsieh}}, \
  and\ \bibinfo {author} {\bibfnamefont {M.}~\bibnamefont {Oshikawa}},\
  }\href@noop {} {\bibfield  {journal} {\bibinfo  {journal} {Phys. Rev. Lett.}\
  }\textbf {\bibinfo {volume} {123}},\ \bibinfo {pages} {180201} (\bibinfo
  {year} {2019})}\BibitemShut {NoStop}%
\bibitem [{\citenamefont {Chen}\ \emph {et~al.}(2011)\citenamefont {Chen},
  \citenamefont {Gu},\ and\ \citenamefont
  {Wen}}]{Chen-Gu-Wen_classification2010}%
  \BibitemOpen
  \bibfield  {author} {\bibinfo {author} {\bibfnamefont {X.}~\bibnamefont
  {Chen}}, \bibinfo {author} {\bibfnamefont {Z.-C.}\ \bibnamefont {Gu}}, \ and\
  \bibinfo {author} {\bibfnamefont {X.-G.}\ \bibnamefont {Wen}},\ }\href
  {\doibase 10.1103/PhysRevB.83.035107} {\bibfield  {journal} {\bibinfo
  {journal} {Phys. Rev. B}\ }\textbf {\bibinfo {volume} {83}},\ \bibinfo
  {pages} {035107} (\bibinfo {year} {2011})}\BibitemShut {NoStop}%
\bibitem [{\citenamefont {Fuji}(2016)}]{Fuji-SymmetryProtection-PRB2016}%
  \BibitemOpen
  \bibfield  {author} {\bibinfo {author} {\bibfnamefont {Y.}~\bibnamefont
  {Fuji}},\ }\href {\doibase 10.1103/PhysRevB.93.104425} {\bibfield  {journal}
  {\bibinfo  {journal} {Phys. Rev. B}\ }\textbf {\bibinfo {volume} {93}},\
  \bibinfo {pages} {104425} (\bibinfo {year} {2016})}\BibitemShut {NoStop}%
\bibitem [{\citenamefont {Watanabe}\ \emph {et~al.}(2015)\citenamefont
  {Watanabe}, \citenamefont {Po}, \citenamefont {Vishwanath},\ and\
  \citenamefont {Zaletel}}]{Watanabe:2015aa}%
  \BibitemOpen
  \bibfield  {author} {\bibinfo {author} {\bibfnamefont {H.}~\bibnamefont
  {Watanabe}}, \bibinfo {author} {\bibfnamefont {H.~C.}\ \bibnamefont {Po}},
  \bibinfo {author} {\bibfnamefont {A.}~\bibnamefont {Vishwanath}}, \ and\
  \bibinfo {author} {\bibfnamefont {M.}~\bibnamefont {Zaletel}},\ }\href@noop
  {} {\bibfield  {journal} {\bibinfo  {journal} {Proc. Natl. Acad. Sci. USA}\
  }\textbf {\bibinfo {volume} {112}},\ \bibinfo {pages} {14551} (\bibinfo
  {year} {2015})}\BibitemShut {NoStop}%
\bibitem [{\citenamefont {Ogata}\ and\ \citenamefont
  {Tasaki}(2018)}]{Ogata:2018aa}%
  \BibitemOpen
  \bibfield  {author} {\bibinfo {author} {\bibfnamefont {Y.}~\bibnamefont
  {Ogata}}\ and\ \bibinfo {author} {\bibfnamefont {H.}~\bibnamefont {Tasaki}},\
  }\href@noop {} {\bibfield  {journal} {\bibinfo  {journal} {arXiv preprint
  arXiv:1808.08740}\ } (\bibinfo {year} {2018})}\BibitemShut {NoStop}%
\bibitem [{\citenamefont {Ogata}\ \emph {et~al.}(2020)\citenamefont {Ogata},
  \citenamefont {Tachikawa},\ and\ \citenamefont {Tasaki}}]{Ogata:2020aa}%
  \BibitemOpen
  \bibfield  {author} {\bibinfo {author} {\bibfnamefont {Y.}~\bibnamefont
  {Ogata}}, \bibinfo {author} {\bibfnamefont {Y.}~\bibnamefont {Tachikawa}}, \
  and\ \bibinfo {author} {\bibfnamefont {H.}~\bibnamefont {Tasaki}},\
  }\href@noop {} {\bibfield  {journal} {\bibinfo  {journal} {arXiv preprint
  arXiv:2004.06458}\ } (\bibinfo {year} {2020})}\BibitemShut {NoStop}%
\bibitem [{\citenamefont {Po}\ \emph {et~al.}(2017)\citenamefont {Po},
  \citenamefont {Watanabe}, \citenamefont {Jian},\ and\ \citenamefont
  {Zaletel}}]{Po:2017aa}%
  \BibitemOpen
  \bibfield  {author} {\bibinfo {author} {\bibfnamefont {H.~C.}\ \bibnamefont
  {Po}}, \bibinfo {author} {\bibfnamefont {H.}~\bibnamefont {Watanabe}},
  \bibinfo {author} {\bibfnamefont {C.-M.}\ \bibnamefont {Jian}}, \ and\
  \bibinfo {author} {\bibfnamefont {M.~P.}\ \bibnamefont {Zaletel}},\
  }\href@noop {} {\bibfield  {journal} {\bibinfo  {journal} {Phys. Rev. Lett.}\
  }\textbf {\bibinfo {volume} {119}},\ \bibinfo {pages} {127202} (\bibinfo
  {year} {2017})}\BibitemShut {NoStop}%
\bibitem [{\citenamefont {Else}\ and\ \citenamefont
  {Thorngren}(2020)}]{Else:2020aa}%
  \BibitemOpen
  \bibfield  {author} {\bibinfo {author} {\bibfnamefont {D.~V.}\ \bibnamefont
  {Else}}\ and\ \bibinfo {author} {\bibfnamefont {R.}~\bibnamefont
  {Thorngren}},\ }\href@noop {} {\bibfield  {journal} {\bibinfo  {journal}
  {Phys. Rev. B}\ }\textbf {\bibinfo {volume} {101}},\ \bibinfo {pages}
  {224437} (\bibinfo {year} {2020})}\BibitemShut {NoStop}%
\bibitem [{\citenamefont {Cheng}\ \emph {et~al.}(2016)\citenamefont {Cheng},
  \citenamefont {Zaletel}, \citenamefont {Barkeshli}, \citenamefont
  {Vishwanath},\ and\ \citenamefont {Bonderson}}]{Cheng:2016aa}%
  \BibitemOpen
  \bibfield  {author} {\bibinfo {author} {\bibfnamefont {M.}~\bibnamefont
  {Cheng}}, \bibinfo {author} {\bibfnamefont {M.}~\bibnamefont {Zaletel}},
  \bibinfo {author} {\bibfnamefont {M.}~\bibnamefont {Barkeshli}}, \bibinfo
  {author} {\bibfnamefont {A.}~\bibnamefont {Vishwanath}}, \ and\ \bibinfo
  {author} {\bibfnamefont {P.}~\bibnamefont {Bonderson}},\ }\href {\doibase
  10.1103/PhysRevX.6.041068} {\bibfield  {journal} {\bibinfo  {journal} {Phys.
  Rev. X}\ }\textbf {\bibinfo {volume} {6}},\ \bibinfo {pages} {041068}
  (\bibinfo {year} {2016})}\BibitemShut {NoStop}%
\bibitem [{Note1()}]{Note1}%
  \BibitemOpen
  \bibinfo {note} {The argument based on the Schmidt decomposition in
  Ref.~\cite {Watanabe:2015aa} is essentially one-dimensional. Its validity in
  higher dimensions may be questionable, especially since there are
  symmetry-protected topological (SPT) phases which have a unique gapped ground
  state with gapless entanglement spectrum, such as the $S=2$
  Affleck-Kennedy-Lieb-Tasaki state~\cite {CPSV-EntSpec_PRB2011}. The other
  argument in Ref.~\cite {Watanabe:2015aa} is based on the observation that (in
  the simplest case) the system containing an odd number of spins has exactly
  degenerate ground states. While it is interesting, by itself it would not be
  sufficient to establish the statement, as the degeneracy is rather trivial.
  Both arguments also depends on the special choice of the system size, which
  we can avoid by using the tilted boundary condition. The arguments in later
  works~\cite {Cheng:2016aa,Po:2017aa,Else:2020aa} either depend on Ref.~\cite
  {Watanabe:2015aa}, or are abstract ones not directly on concrete lattice
  models.}\BibitemShut {Stop}%
\bibitem [{\citenamefont {Hirano}\ \emph {et~al.}(2008)\citenamefont {Hirano},
  \citenamefont {Katsura},\ and\ \citenamefont {Hatsugai}}]{Hirano:2008aa}%
  \BibitemOpen
  \bibfield  {author} {\bibinfo {author} {\bibfnamefont {T.}~\bibnamefont
  {Hirano}}, \bibinfo {author} {\bibfnamefont {H.}~\bibnamefont {Katsura}}, \
  and\ \bibinfo {author} {\bibfnamefont {Y.}~\bibnamefont {Hatsugai}},\
  }\href@noop {} {\bibfield  {journal} {\bibinfo  {journal} {Phys. Rev. B}\
  }\textbf {\bibinfo {volume} {78}},\ \bibinfo {pages} {054431} (\bibinfo
  {year} {2008})}\BibitemShut {NoStop}%
\bibitem [{\citenamefont {Yao}\ and\ \citenamefont
  {Oshikawa}(2020)}]{Yao:2020PRX}%
  \BibitemOpen
  \bibfield  {author} {\bibinfo {author} {\bibfnamefont {Y.}~\bibnamefont
  {Yao}}\ and\ \bibinfo {author} {\bibfnamefont {M.}~\bibnamefont {Oshikawa}},\
  }\href {\doibase 10.1103/PhysRevX.10.031008} {\bibfield  {journal} {\bibinfo
  {journal} {Phys. Rev. X}\ }\textbf {\bibinfo {volume} {10}},\ \bibinfo
  {pages} {031008} (\bibinfo {year} {2020})}\BibitemShut {NoStop}%
\bibitem [{Note2()}]{Note2}%
  \BibitemOpen
  \bibinfo {note} {See Supplemental Materials for a diagrammatic proof and the
  detailed definition of $\protect \tilde {T}_1$.}\BibitemShut {Stop}%
\bibitem [{\citenamefont {Watanabe}(2018)}]{Watanabe:2018aa}%
  \BibitemOpen
  \bibfield  {author} {\bibinfo {author} {\bibfnamefont {H.}~\bibnamefont
  {Watanabe}},\ }\href@noop {} {\bibfield  {journal} {\bibinfo  {journal}
  {Phys. Rev. B}\ }\textbf {\bibinfo {volume} {98}},\ \bibinfo {pages} {155137}
  (\bibinfo {year} {2018})}\BibitemShut {NoStop}%
\bibitem [{\citenamefont {Hastings}\ and\ \citenamefont
  {Koma}(2006)}]{Hastings:2006aa}%
  \BibitemOpen
  \bibfield  {author} {\bibinfo {author} {\bibfnamefont {M.~B.}\ \bibnamefont
  {Hastings}}\ and\ \bibinfo {author} {\bibfnamefont {T.}~\bibnamefont
  {Koma}},\ }\href {\doibase 10.1007/s00220-006-0030-4} {\bibfield  {journal}
  {\bibinfo  {journal} {Commun. Math. Phys.}\ }\textbf {\bibinfo {volume}
  {265}},\ \bibinfo {pages} {781} (\bibinfo {year} {2006})}\BibitemShut
  {NoStop}%
\bibitem [{\citenamefont {Betsuyaku}(1984)}]{Betsuyaku-PRL1984}%
  \BibitemOpen
  \bibfield  {author} {\bibinfo {author} {\bibfnamefont {H.}~\bibnamefont
  {Betsuyaku}},\ }\href {\doibase 10.1103/PhysRevLett.53.629} {\bibfield
  {journal} {\bibinfo  {journal} {Phys. Rev. Lett.}\ }\textbf {\bibinfo
  {volume} {53}},\ \bibinfo {pages} {629} (\bibinfo {year} {1984})}\BibitemShut
  {NoStop}%
\bibitem [{\citenamefont {Piroli}\ \emph {et~al.}(2017)\citenamefont {Piroli},
  \citenamefont {Pozsgay},\ and\ \citenamefont {Vernier}}]{Piroli-QTM2017}%
  \BibitemOpen
  \bibfield  {author} {\bibinfo {author} {\bibfnamefont {L.}~\bibnamefont
  {Piroli}}, \bibinfo {author} {\bibfnamefont {B.}~\bibnamefont {Pozsgay}}, \
  and\ \bibinfo {author} {\bibfnamefont {E.}~\bibnamefont {Vernier}},\ }\href
  {\doibase 10.1088/1742-5468/aa5d1e} {\bibfield  {journal} {\bibinfo
  {journal} {Journal of Statistical Mechanics: Theory and Experiment}\ }\textbf
  {\bibinfo {volume} {2017}},\ \bibinfo {pages} {023106} (\bibinfo {year}
  {2017})}\BibitemShut {NoStop}%
\bibitem [{\citenamefont {Suzuki}(1976)}]{Suzuki:1976aa}%
  \BibitemOpen
  \bibfield  {author} {\bibinfo {author} {\bibfnamefont {M.}~\bibnamefont
  {Suzuki}},\ }\href@noop {} {\bibfield  {journal} {\bibinfo  {journal} {Progr.
  Theor. Phys.}\ }\textbf {\bibinfo {volume} {56}},\ \bibinfo {pages} {1454}
  (\bibinfo {year} {1976})}\BibitemShut {NoStop}%
\bibitem [{Note3()}]{Note3}%
  \BibitemOpen
  \bibinfo {note} {The mapping to the classical statistical system is exact
  only in the limit of an infinite Trotter number. Here we assume that the QTM
  is well-defined in this limit.}\BibitemShut {Stop}%
\bibitem [{\citenamefont {Crosswhite}\ and\ \citenamefont
  {Bacon}(2008)}]{CrossWhite-Bacon-MPO_PRA2008}%
  \BibitemOpen
  \bibfield  {author} {\bibinfo {author} {\bibfnamefont {G.~M.}\ \bibnamefont
  {Crosswhite}}\ and\ \bibinfo {author} {\bibfnamefont {D.}~\bibnamefont
  {Bacon}},\ }\href {\doibase 10.1103/PhysRevA.78.012356} {\bibfield  {journal}
  {\bibinfo  {journal} {Phys. Rev. A}\ }\textbf {\bibinfo {volume} {78}},\
  \bibinfo {pages} {012356} (\bibinfo {year} {2008})}\BibitemShut {NoStop}%
\bibitem [{\citenamefont {Pirvu}\ \emph {et~al.}(2010)\citenamefont {Pirvu},
  \citenamefont {Murg}, \citenamefont {Cirac},\ and\ \citenamefont
  {Verstraete}}]{Pirvu-MPO_NJP2010}%
  \BibitemOpen
  \bibfield  {author} {\bibinfo {author} {\bibfnamefont {B.}~\bibnamefont
  {Pirvu}}, \bibinfo {author} {\bibfnamefont {V.}~\bibnamefont {Murg}},
  \bibinfo {author} {\bibfnamefont {J.~I.}\ \bibnamefont {Cirac}}, \ and\
  \bibinfo {author} {\bibfnamefont {F.}~\bibnamefont {Verstraete}},\ }\href
  {\doibase 10.1088/1367-2630/12/2/025012} {\bibfield  {journal} {\bibinfo
  {journal} {New Journal of Physics}\ }\textbf {\bibinfo {volume} {12}},\
  \bibinfo {pages} {025012} (\bibinfo {year} {2010})}\BibitemShut {NoStop}%
\bibitem [{Note4()}]{Note4}%
  \BibitemOpen
  \bibinfo {note} {See Supplemental Material for more details.}\BibitemShut
  {Stop}%
\bibitem [{Note5()}]{Note5}%
  \BibitemOpen
  \bibinfo {note} {See Supplemental Material for the proofs.}\BibitemShut
  {Stop}%
\bibitem [{\citenamefont {Kishimoto}\ \emph {et~al.}(2018)\citenamefont
  {Kishimoto}, \citenamefont {Morita}, \citenamefont {Matsubayashi},
  \citenamefont {Sota}, \citenamefont {Yunoki},\ and\ \citenamefont
  {Tohyama}}]{Kishimoto:2018aa}%
  \BibitemOpen
  \bibfield  {author} {\bibinfo {author} {\bibfnamefont {M.}~\bibnamefont
  {Kishimoto}}, \bibinfo {author} {\bibfnamefont {K.}~\bibnamefont {Morita}},
  \bibinfo {author} {\bibfnamefont {Y.}~\bibnamefont {Matsubayashi}}, \bibinfo
  {author} {\bibfnamefont {S.}~\bibnamefont {Sota}}, \bibinfo {author}
  {\bibfnamefont {S.}~\bibnamefont {Yunoki}}, \ and\ \bibinfo {author}
  {\bibfnamefont {T.}~\bibnamefont {Tohyama}},\ }\href@noop {} {\bibfield
  {journal} {\bibinfo  {journal} {Phys. Rev. B}\ }\textbf {\bibinfo {volume}
  {98}},\ \bibinfo {pages} {054411} (\bibinfo {year} {2018})}\BibitemShut
  {NoStop}%
\bibitem [{\citenamefont {Maksimov}\ \emph {et~al.}(2019)\citenamefont
  {Maksimov}, \citenamefont {Zhu}, \citenamefont {White},\ and\ \citenamefont
  {Chernyshev}}]{Maksimov:2019aa}%
  \BibitemOpen
  \bibfield  {author} {\bibinfo {author} {\bibfnamefont {P.~A.}~\bibnamefont
  {Maksimov}}, \bibinfo {author} {\bibfnamefont {Z.}~\bibnamefont {Zhu}},
  \bibinfo {author} {\bibfnamefont {S.~R.}\ \bibnamefont {White}}, \ and\
  \bibinfo {author} {\bibfnamefont {A.~L.}~\bibnamefont {Chernyshev}},\
  }\href@noop {} {\bibfield  {journal} {\bibinfo  {journal} {Phys. Rev. X}\
  }\textbf {\bibinfo {volume} {9}},\ \bibinfo {pages} {021017} (\bibinfo {year}
  {2019})}\BibitemShut {NoStop}%
\bibitem [{\citenamefont {Kitaev}(2006)}]{Kitaev:2006aa}%
  \BibitemOpen
  \bibfield  {author} {\bibinfo {author} {\bibfnamefont {A.}~\bibnamefont
  {Kitaev}},\ }\href@noop {} {\bibfield  {journal} {\bibinfo  {journal} {Ann.
  Phys.}\ }\textbf {\bibinfo {volume} {321}},\ \bibinfo {pages} {2} (\bibinfo
  {year} {2006})}\BibitemShut {NoStop}%
\bibitem [{\citenamefont {Kimchi}\ \emph {et~al.}(2013)\citenamefont {Kimchi},
  \citenamefont {Parameswaran}, \citenamefont {Turner}, \citenamefont {Wang},\
  and\ \citenamefont {Vishwanath}}]{Kimchi-featureless-honeycomb}%
  \BibitemOpen
  \bibfield  {author} {\bibinfo {author} {\bibfnamefont {I.}~\bibnamefont
  {Kimchi}}, \bibinfo {author} {\bibfnamefont {S.~A.}\ \bibnamefont
  {Parameswaran}}, \bibinfo {author} {\bibfnamefont {A.~M.}\ \bibnamefont
  {Turner}}, \bibinfo {author} {\bibfnamefont {F.}~\bibnamefont {Wang}}, \ and\
  \bibinfo {author} {\bibfnamefont {A.}~\bibnamefont {Vishwanath}},\ }\href
  {\doibase 10.1073/pnas.1307245110} {\bibfield  {journal} {\bibinfo  {journal}
  {Proceedings of the National Academy of Sciences}\ }\textbf {\bibinfo
  {volume} {110}},\ \bibinfo {pages} {16378} (\bibinfo {year} {2013})},\
  \Eprint
  {http://arxiv.org/abs/https://www.pnas.org/content/110/41/16378.full.pdf}
  {https://www.pnas.org/content/110/41/16378.full.pdf} \BibitemShut {NoStop}%
\bibitem [{\citenamefont {Schweigert}(1997)}]{Schweigert:1997aa}%
  \BibitemOpen
  \bibfield  {author} {\bibinfo {author} {\bibfnamefont {C.}~\bibnamefont
  {Schweigert}},\ }\href@noop {} {\bibfield  {journal} {\bibinfo  {journal}
  {Nucl. Phys. B}\ }\textbf {\bibinfo {volume} {492}},\ \bibinfo {pages} {743}
  (\bibinfo {year} {1997})}\BibitemShut {NoStop}%
\bibitem [{\citenamefont {Borel}\ \emph {et~al.}(2002)\citenamefont {Borel},
  \citenamefont {Friedman}, \citenamefont {Morgan},\ and\ \citenamefont
  {Morgan}}]{Borel:2002aa}%
  \BibitemOpen
  \bibfield  {author} {\bibinfo {author} {\bibfnamefont {A.}~\bibnamefont
  {Borel}}, \bibinfo {author} {\bibfnamefont {R.}~\bibnamefont {Friedman}},
  \bibinfo {author} {\bibfnamefont {J.~W.}\ \bibnamefont {Morgan}}, \ and\
  \bibinfo {author} {\bibfnamefont {J.~W.}\ \bibnamefont {Morgan}},\
  }\href@noop {} {\emph {\bibinfo {title} {Almost commuting elements in compact
  Lie groups}}}\ (\bibinfo  {publisher} {American Mathematical Soc.},\ \bibinfo
  {year} {2002})\BibitemShut {NoStop}%
\bibitem [{\citenamefont {Kac}\ and\ \citenamefont
  {Smilga}(2000)}]{Kac:2000aa}%
  \BibitemOpen
  \bibfield  {author} {\bibinfo {author} {\bibfnamefont {V.~G.}\ \bibnamefont
  {Kac}}\ and\ \bibinfo {author} {\bibfnamefont {A.~V.}\ \bibnamefont
  {Smilga}},\ }\enquote {\bibinfo {title} {Vacuum structure in supersymmetric
  yang--mills theories with any gauge group},}\ in\ \href@noop {} {\emph
  {\bibinfo {booktitle} {The Many Faces of the Superworld: Yuri Golfand
  Memorial Volume}}}\ (\bibinfo  {publisher} {World Scientific},\ \bibinfo
  {year} {2000})\ pp.\ \bibinfo {pages} {185--234}\BibitemShut {NoStop}%
\bibitem [{\citenamefont {Georgi}(2000)}]{Georgi:2000}%
  \BibitemOpen
  \bibfield  {author} {\bibinfo {author} {\bibfnamefont {H.}~\bibnamefont
  {Georgi}},\ }\href@noop {} {\emph {\bibinfo {title} {Lie Algebras In Particle
  Physics}}}\ (\bibinfo  {publisher} {Boca Raton: CRC Press},\ \bibinfo {year}
  {2000})\BibitemShut {NoStop}%
\bibitem [{\citenamefont {Affleck}\ and\ \citenamefont
  {Haldane}(1987)}]{Affleck:1987aa}%
  \BibitemOpen
  \bibfield  {author} {\bibinfo {author} {\bibfnamefont {I.}~\bibnamefont
  {Affleck}}\ and\ \bibinfo {author} {\bibfnamefont {F.~D.~M.}\ \bibnamefont
  {Haldane}},\ }\href {https://link.aps.org/doi/10.1103/PhysRevB.36.5291}
  {\bibfield  {journal} {\bibinfo  {journal} {Phys. Rev. B}\ }\textbf {\bibinfo
  {volume} {36}},\ \bibinfo {pages} {5291} (\bibinfo {year}
  {1987})}\BibitemShut {NoStop}%
\bibitem [{\citenamefont {Affleck}(1988)}]{Affleck:1988aa}%
  \BibitemOpen
  \bibfield  {author} {\bibinfo {author} {\bibfnamefont {I.}~\bibnamefont
  {Affleck}},\ }\href@noop {} {\bibfield  {journal} {\bibinfo  {journal} {Nucl.
  Phys. B}\ }\textbf {\bibinfo {volume} {305}},\ \bibinfo {pages} {582}
  (\bibinfo {year} {1988})}\BibitemShut {NoStop}%
\bibitem [{\citenamefont {Wu}\ \emph {et~al.}(2003)\citenamefont {Wu},
  \citenamefont {Hu},\ and\ \citenamefont {Zhang}}]{Wu:2003aa}%
  \BibitemOpen
  \bibfield  {author} {\bibinfo {author} {\bibfnamefont {C.}~\bibnamefont
  {Wu}}, \bibinfo {author} {\bibfnamefont {J.-P.}\ \bibnamefont {Hu}}, \ and\
  \bibinfo {author} {\bibfnamefont {S.-C.}\ \bibnamefont {Zhang}},\ }\href
  {https://link.aps.org/doi/10.1103/PhysRevLett.91.186402} {\bibfield
  {journal} {\bibinfo  {journal} {Phys. Rev. Lett.}\ }\textbf {\bibinfo
  {volume} {91}},\ \bibinfo {pages} {186402} (\bibinfo {year}
  {2003})}\BibitemShut {NoStop}%
\bibitem [{\citenamefont {Honerkamp}\ and\ \citenamefont
  {Hofstetter}(2004)}]{Honerkamp:2004aa}%
  \BibitemOpen
  \bibfield  {author} {\bibinfo {author} {\bibfnamefont {C.}~\bibnamefont
  {Honerkamp}}\ and\ \bibinfo {author} {\bibfnamefont {W.}~\bibnamefont
  {Hofstetter}},\ }\href
  {https://link.aps.org/doi/10.1103/PhysRevLett.92.170403} {\bibfield
  {journal} {\bibinfo  {journal} {Phys. Rev. Lett.}\ }\textbf {\bibinfo
  {volume} {92}},\ \bibinfo {pages} {170403} (\bibinfo {year}
  {2004})}\BibitemShut {NoStop}%
\bibitem [{\citenamefont {Cazalilla}\ \emph {et~al.}(2009)\citenamefont
  {Cazalilla}, \citenamefont {Ho},\ and\ \citenamefont
  {Ueda}}]{Cazalilla:2009aa}%
  \BibitemOpen
  \bibfield  {author} {\bibinfo {author} {\bibfnamefont {M.~A.}\ \bibnamefont
  {Cazalilla}}, \bibinfo {author} {\bibfnamefont {A.}~\bibnamefont {Ho}}, \
  and\ \bibinfo {author} {\bibfnamefont {M.}~\bibnamefont {Ueda}},\ }\href@noop
  {} {\bibfield  {journal} {\bibinfo  {journal} {New J. Phys.}\ }\textbf
  {\bibinfo {volume} {11}} (\bibinfo {year} {2009})}\BibitemShut {NoStop}%
\bibitem [{\citenamefont {Gorshkov}\ \emph {et~al.}(2010)\citenamefont
  {Gorshkov}, \citenamefont {Hermele}, \citenamefont {Gurarie}, \citenamefont
  {Xu}, \citenamefont {Julienne}, \citenamefont {Ye}, \citenamefont {Zoller},
  \citenamefont {Demler}, \citenamefont {Lukin},\ and\ \citenamefont
  {Rey}}]{Gorshkov:2010aa}%
  \BibitemOpen
  \bibfield  {author} {\bibinfo {author} {\bibfnamefont {A.~V.}\ \bibnamefont
  {Gorshkov}}, \bibinfo {author} {\bibfnamefont {M.}~\bibnamefont {Hermele}},
  \bibinfo {author} {\bibfnamefont {V.}~\bibnamefont {Gurarie}}, \bibinfo
  {author} {\bibfnamefont {C.}~\bibnamefont {Xu}}, \bibinfo {author}
  {\bibfnamefont {P.~S.}\ \bibnamefont {Julienne}}, \bibinfo {author}
  {\bibfnamefont {J.}~\bibnamefont {Ye}}, \bibinfo {author} {\bibfnamefont
  {P.}~\bibnamefont {Zoller}}, \bibinfo {author} {\bibfnamefont
  {E.}~\bibnamefont {Demler}}, \bibinfo {author} {\bibfnamefont {M.~D.}\
  \bibnamefont {Lukin}}, \ and\ \bibinfo {author} {\bibfnamefont
  {A.}~\bibnamefont {Rey}},\ }\href@noop {} {\bibfield  {journal} {\bibinfo
  {journal} {Nature physics}\ }\textbf {\bibinfo {volume} {6}},\ \bibinfo
  {pages} {289} (\bibinfo {year} {2010})}\BibitemShut {NoStop}%
\bibitem [{\citenamefont {Taie}\ \emph {et~al.}(2012)\citenamefont {Taie},
  \citenamefont {Yamazaki}, \citenamefont {Sugawa},\ and\ \citenamefont
  {Takahashi}}]{Taie:2012aa}%
  \BibitemOpen
  \bibfield  {author} {\bibinfo {author} {\bibfnamefont {S.}~\bibnamefont
  {Taie}}, \bibinfo {author} {\bibfnamefont {R.}~\bibnamefont {Yamazaki}},
  \bibinfo {author} {\bibfnamefont {S.}~\bibnamefont {Sugawa}}, \ and\ \bibinfo
  {author} {\bibfnamefont {Y.}~\bibnamefont {Takahashi}},\ }\href@noop {}
  {\bibfield  {journal} {\bibinfo  {journal} {Nature Physics}\ }\textbf
  {\bibinfo {volume} {8}},\ \bibinfo {pages} {825} (\bibinfo {year}
  {2012})}\BibitemShut {NoStop}%
\bibitem [{\citenamefont {Pagano}\ \emph {et~al.}(2014)\citenamefont {Pagano},
  \citenamefont {Mancini}, \citenamefont {Cappellini}, \citenamefont
  {Lombardi}, \citenamefont {Sch{\"a}fer}, \citenamefont {Hu}, \citenamefont
  {Liu}, \citenamefont {Catani}, \citenamefont {Sias},\ and\ \citenamefont
  {Inguscio}}]{Pagano:2014aa}%
  \BibitemOpen
  \bibfield  {author} {\bibinfo {author} {\bibfnamefont {G.}~\bibnamefont
  {Pagano}}, \bibinfo {author} {\bibfnamefont {M.}~\bibnamefont {Mancini}},
  \bibinfo {author} {\bibfnamefont {G.}~\bibnamefont {Cappellini}}, \bibinfo
  {author} {\bibfnamefont {P.}~\bibnamefont {Lombardi}}, \bibinfo {author}
  {\bibfnamefont {F.}~\bibnamefont {Sch{\"a}fer}}, \bibinfo {author}
  {\bibfnamefont {H.}~\bibnamefont {Hu}}, \bibinfo {author} {\bibfnamefont
  {X.-J.}\ \bibnamefont {Liu}}, \bibinfo {author} {\bibfnamefont
  {J.}~\bibnamefont {Catani}}, \bibinfo {author} {\bibfnamefont
  {C.}~\bibnamefont {Sias}}, \ and\ \bibinfo {author} {\bibfnamefont
  {M.}~\bibnamefont {Inguscio}},\ }\href@noop {} {\bibfield  {journal}
  {\bibinfo  {journal} {Nature Physics}\ }\textbf {\bibinfo {volume} {10}},\
  \bibinfo {pages} {198} (\bibinfo {year} {2014})}\BibitemShut {NoStop}%
\bibitem [{\citenamefont {Scazza}\ \emph {et~al.}(2014)\citenamefont {Scazza},
  \citenamefont {Hofrichter}, \citenamefont {H{\"o}fer}, \citenamefont
  {De~Groot}, \citenamefont {Bloch},\ and\ \citenamefont
  {F{\"o}lling}}]{Scazza:2014aa}%
  \BibitemOpen
  \bibfield  {author} {\bibinfo {author} {\bibfnamefont {F.}~\bibnamefont
  {Scazza}}, \bibinfo {author} {\bibfnamefont {C.}~\bibnamefont {Hofrichter}},
  \bibinfo {author} {\bibfnamefont {M.}~\bibnamefont {H{\"o}fer}}, \bibinfo
  {author} {\bibfnamefont {P.}~\bibnamefont {De~Groot}}, \bibinfo {author}
  {\bibfnamefont {I.}~\bibnamefont {Bloch}}, \ and\ \bibinfo {author}
  {\bibfnamefont {S.}~\bibnamefont {F{\"o}lling}},\ }\href@noop {} {\bibfield
  {journal} {\bibinfo  {journal} {Nature Physics}\ }\textbf {\bibinfo {volume}
  {10}},\ \bibinfo {pages} {779} (\bibinfo {year} {2014})}\BibitemShut
  {NoStop}%
\bibitem [{\citenamefont {Zhang}\ \emph {et~al.}(2014)\citenamefont {Zhang},
  \citenamefont {Bishof}, \citenamefont {Bromley}, \citenamefont {Kraus},
  \citenamefont {Safronova}, \citenamefont {Zoller}, \citenamefont {Rey},\ and\
  \citenamefont {Ye}}]{Zhang:2014aa}%
  \BibitemOpen
  \bibfield  {author} {\bibinfo {author} {\bibfnamefont {X.}~\bibnamefont
  {Zhang}}, \bibinfo {author} {\bibfnamefont {M.}~\bibnamefont {Bishof}},
  \bibinfo {author} {\bibfnamefont {S.}~\bibnamefont {Bromley}}, \bibinfo
  {author} {\bibfnamefont {C.}~\bibnamefont {Kraus}}, \bibinfo {author}
  {\bibfnamefont {M.}~\bibnamefont {Safronova}}, \bibinfo {author}
  {\bibfnamefont {P.}~\bibnamefont {Zoller}}, \bibinfo {author} {\bibfnamefont
  {A.~M.}\ \bibnamefont {Rey}}, \ and\ \bibinfo {author} {\bibfnamefont
  {J.}~\bibnamefont {Ye}},\ }\href@noop {} {\bibfield  {journal} {\bibinfo
  {journal} {science}\ }\textbf {\bibinfo {volume} {345}},\ \bibinfo {pages}
  {1467} (\bibinfo {year} {2014})}\BibitemShut {NoStop}%
\bibitem [{Note6()}]{Note6}%
  \BibitemOpen
  \bibinfo {note} {See Supplemental Materials for an explicit form of $V_N$ and
  $W_N$, and a generalized $XYZ$ model.}\BibitemShut {Stop}%
\bibitem [{\citenamefont {Shimizu}\ and\ \citenamefont
  {Yonekura}(2018)}]{Shimizu:2018aa}%
  \BibitemOpen
  \bibfield  {author} {\bibinfo {author} {\bibfnamefont {H.}~\bibnamefont
  {Shimizu}}\ and\ \bibinfo {author} {\bibfnamefont {K.}~\bibnamefont
  {Yonekura}},\ }\href {\doibase 10.1103/PhysRevD.97.105011} {\bibfield
  {journal} {\bibinfo  {journal} {Phys. Rev. D}\ }\textbf {\bibinfo {volume}
  {97}},\ \bibinfo {pages} {105011} (\bibinfo {year} {2018})}\BibitemShut
  {NoStop}%
\bibitem [{\citenamefont {Tanizaki}\ and\ \citenamefont
  {Kikuchi}(2017)}]{Tanizaki:2017aa}%
  \BibitemOpen
  \bibfield  {author} {\bibinfo {author} {\bibfnamefont {Y.}~\bibnamefont
  {Tanizaki}}\ and\ \bibinfo {author} {\bibfnamefont {Y.}~\bibnamefont
  {Kikuchi}},\ }\href {\doibase 10.1007/JHEP06(2017)102} {\bibfield  {journal}
  {\bibinfo  {journal} {JHEP}\ }\textbf {\bibinfo {volume} {2017}},\ \bibinfo
  {pages} {102} (\bibinfo {year} {2017})}\BibitemShut {NoStop}%
\bibitem [{\citenamefont {Cirac}\ \emph {et~al.}(2011)\citenamefont {Cirac},
  \citenamefont {Poilblanc}, \citenamefont {Schuch},\ and\ \citenamefont
  {Verstraete}}]{CPSV-EntSpec_PRB2011}%
  \BibitemOpen
  \bibfield  {author} {\bibinfo {author} {\bibfnamefont {J.~I.}\ \bibnamefont
  {Cirac}}, \bibinfo {author} {\bibfnamefont {D.}~\bibnamefont {Poilblanc}},
  \bibinfo {author} {\bibfnamefont {N.}~\bibnamefont {Schuch}}, \ and\ \bibinfo
  {author} {\bibfnamefont {F.}~\bibnamefont {Verstraete}},\ }\href {\doibase
  10.1103/PhysRevB.83.245134} {\bibfield  {journal} {\bibinfo  {journal} {Phys.
  Rev. B}\ }\textbf {\bibinfo {volume} {83}},\ \bibinfo {pages} {245134}
  (\bibinfo {year} {2011})}\BibitemShut {NoStop}%
\end{thebibliography}
%merlin.mbs apsrev4-1.bst 2010-07-25 4.21a (PWD, AO, DPC) hacked
%Control: key (0)
%Control: author (8) initials jnrlst
%Control: editor formatted (1) identically to author
%Control: production of article title (-1) disabled
%Control: page (0) single
%Control: year (1) truncated
%Control: production of eprint (0) enabled
%
\newpage
\begin{widetext}
\appendix
\section{An explicit form of $\mathbb{Z}_N\times\mathbb{Z}_N\subset\text{PSU}(N)$ and a generalized $XYZ$ model}
In this part, we explicitly give the operator form of $v_{N,\vec{r}}$ and $w_{N,\vec{r}}$ by SU$(N)$-spin degrees of freedom as
\begin{eqnarray}
v_{N,\vec{r}}&=&\left[\exp\left(i\pi\frac{\mathcal{S}^1_{2,\vec{r}}+\mathcal{S}^2_{1,\vec{r}}}{2}\right)\exp\left(-i\pi\frac{\mathcal{S}^1_{1,\vec{r}}+\mathcal{S}^2_{2,\vec{r}}}{2}\right)\right]\cdots\nonumber\\
&&\cdots\left[\exp\left(i\pi\frac{\mathcal{S}^1_{N,\vec{r}}+\mathcal{S}^N_{1,\vec{r}}}{2}\right)\exp\left(-i\pi\frac{\mathcal{S}^1_{1,\vec{r}}+\mathcal{S}^N_{N,\vec{r}}}{2}\right)\right];\nonumber\\
w_{N,\vec{r}}&=&\exp\left(-\sum_{\alpha}i\frac{2\pi}{N}\alpha \mathcal{S}^\alpha_{\alpha,\vec{r}}\right).
\end{eqnarray}
$V_N=\prod_{\vec{r}}v_{N,\vec{r}}$ and $W_N=\prod_{\vec{r}}w_{N,\vec{r}}$ reduce to $R^\pi_x$ and $R^\pi_z$, respectively, when $N=2$. 
They generate the $\mathbb{Z}_N\times\mathbb{Z}_N\subset\text{PSU}(N)$ in the (adjoint) representation: 
\begin{eqnarray}
\mathcal{S}^\alpha_\beta\rightarrow U^\dagger\mathcal{S}^\alpha_\beta U, 
\end{eqnarray}
where $U$ is any finite product of $V_N$ and $W_N$. 
If the local Hilbert space per unit cell is a tensor product of these Yang-tableaux and we denote the total number of the boxes in this general tensor product by $b$, such as
\begin{eqnarray}
\underbrace{\yng(2,1)\otimes\yng(1)\otimes\yng(2)\otimes\cdots}_\textrm{$b$ YT boxes},
\end{eqnarray}
we obtain
\begin{eqnarray}
v_{N,\vec{r}}w_{N,\vec{r}}=\exp\left(i2\pi\frac{b}{N}\right)w_{N,\vec{r}}v_{N,\vec{r}},
\end{eqnarray}

A SU$(N)$ generalization of SU$(2)$ $XYZ$ chain possessing this $\mathbb{Z}_N\times\mathbb{Z}_N$ takes the form as
\begin{eqnarray}
\mathcal{H}_{N;XYZ}=\sum_{g\in\mathbb{Z}_N\times\mathbb{Z}_N;i}J_g\mathcal{S}_{\beta,i}^\alpha (g)_{\alpha\beta}(g^\dagger)_{\lambda\varphi}S^\lambda_{\varphi,i+1}+\text{h.c.},\nonumber
\end{eqnarray}
with all the dumb indices $\alpha,\beta,\lambda,\varphi$ summed up.
\section{A proof of the commutator $R^\pi_x\tilde{T}_1=(-1)^{2s}\tilde{T}_1R^\pi_x$}
In this part,
we will give a diagrammatic proof for $R^\pi_x\tilde{T}_1=(-1)^{2s}\tilde{T}_1R^\pi_x$ in the main text when $d=2$ for simplicity.
It can be directly generalized to arbitrary dimensions.

Let us first impose the following tilted boundary condition as in the main text:
\begin{eqnarray}
(T_{1})^{-L_1}\vec{S}_{\vec{r}}(T_1)^{L_1}=\vec{S}_{\vec{r}+\hat{x}_{2}}, 
\end{eqnarray}
which is explicitly sketched in FIG.~\ref{gauge}.
\begin{figure}[h]
\centering
\includegraphics[width=8.5cm,pagebox=cropbox,clip]{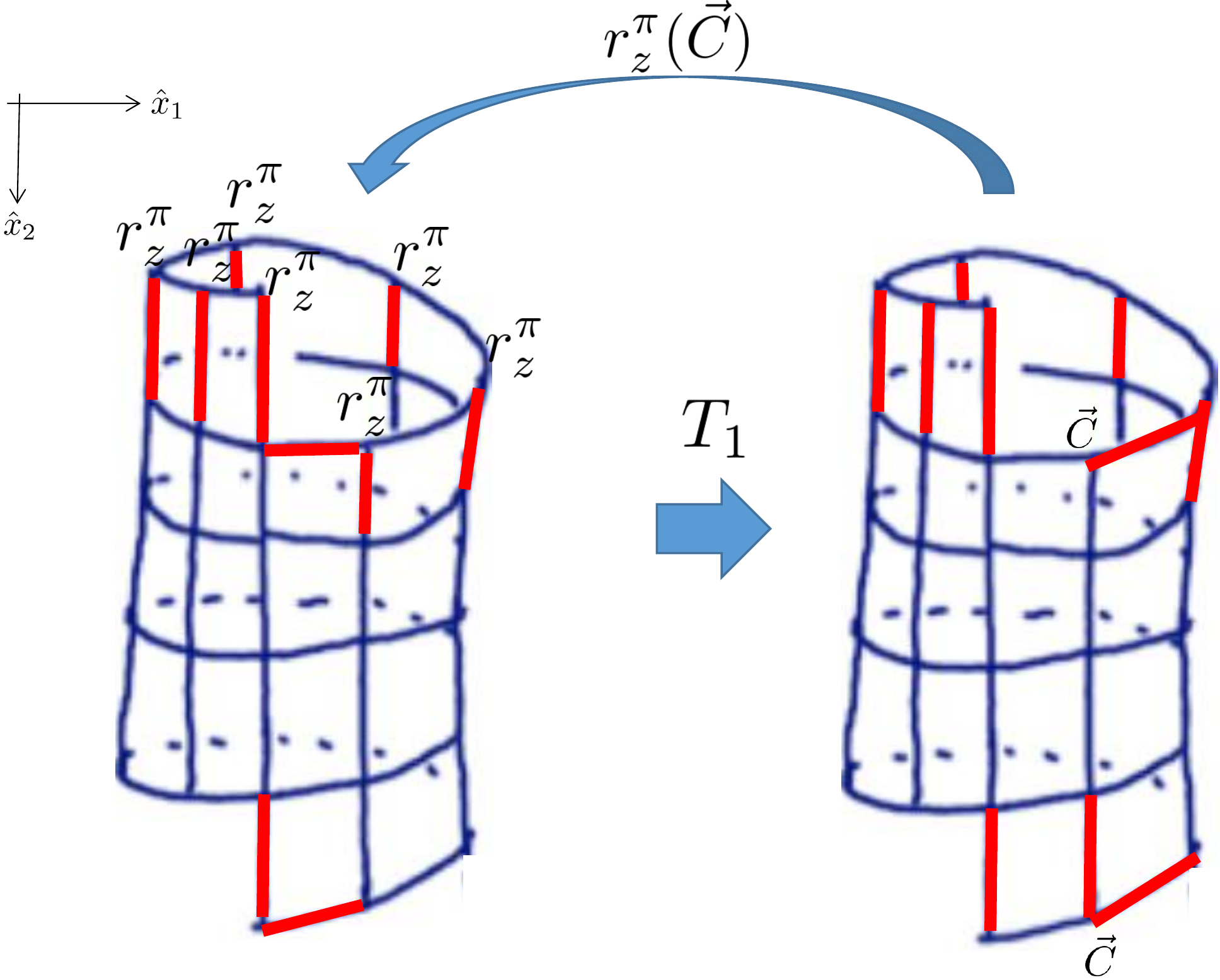}
\caption{The traditional lattice translation $T_1$ is not a symmetry for the twisted Hamiltonian while $\tilde{T}_1\equiv\hat{r}_z^\pi(\vec{C})T_1$ is the proper symmetry to consider.}%
\label{gauge}
\end{figure}

Then we close the lattice to be a lattice torus by imposing the $R^\pi_z$-twisted boundary condition along the $\hat{x}_2$ direction.
For simplicity, let us consider $XYZ$-type nearest-neighbor couplings.
This can be systematically done by twisting the boundary interaction by:
\begin{eqnarray}
H_\text{bd}\rightarrow(R_\text{$z$,bd}^\pi)^{-1}H_\text{bd}R_\text{$z$,bd}^\pi,
\end{eqnarray}
where $H_\text{bd}$ is the interaction terms that crosses the boundary and
\begin{eqnarray}
R_\text{$z$,bd}^\pi\equiv\prod_{\vec{r}\in\{\text{boundary}\}}r_z^\pi(\vec{r}), 
\end{eqnarray}
in which the $r_z^\pi$ actions on those boundary spins are shown in FIG.~\ref{gauge}.
The effect of $R_\text{$z$,bd}^\pi$ on the $XYZ$-type $H_\text{bd}$ is to flip the signs of the terms of $S^zS^z$ while keeping the other $S^{x,y}S^{x,y}$ terms. 
We label these twistings by red bonds on the left side of FIG.~\ref{gauge}.
Two bonds on the bottom are flipped due to the site identification/pasting with the upper side.
It should be noted that there is one (important) horizontal bond near the corner being flipped as well because the site on its right end is a boundary spin which has been acted by $R^\pi_\text{$z$,bd}$ while it, itself, is not a boundary spin.
This horizontal red bond is the high-dimensional analog of the twisted bond in the one-dimensional case. 

Then we do a lattice translation $T_1$ which effectively moves all the red bonds by one lattice site along $\hat{x}_1$.
The Hamiltonian is clearly asymmetric under $T_1$ (partially) due to the movement of the red horizontal bond.
Let us label the site at the left end of the current horizontal bond as $\vec{C}$ as in FIG.~\ref{gauge}.
We can further (adjointly) act the operator $r_z^\pi(\vec{C})$ (so-called ``gauge transformation'' in the main text) on the Hamiltonian so that all the nearest-neighbor bonds connecting $\vec{C}$ will change colors.
Namely, the horizontal red bond on its right and the vertical red bond on the upper side of $\vec{C}$ is annihilated while the bonds on its left and bottom become red. 
The final $r^\pi_z(\vec{C})T_1$-transformed Hamiltonian is nothing but the original Hamiltonian. 
Therefore, the proper translation symmetry to be considered in the twisted tilted boundary condition should be
\begin{eqnarray}
\tilde{T}_1=r^\pi_z(\vec{C})T_1,
\end{eqnarray}
i.e., the lattice translation followed by the gauge transformation. 
Then we could calculate the commutator between $\tilde{T}_1$ and the other symmetry operator
\begin{eqnarray}\label{rx}
R^\pi_x\equiv\prod_{\vec{r}}r^\pi_x(\vec{r})
\end{eqnarray}
as
\begin{eqnarray}
\tilde{T}_1R^\pi_x\tilde{T}_1^{-1}&=&r^\pi_z(\vec{C})(T_1R^\pi_xT_1^{-1})[r^\pi_z(\vec{C})]^{-1}\nonumber\\
&=&r^\pi_z(\vec{C})R^\pi_x[r^\pi_z(\vec{C})]^{-1}\nonumber\\
&=&\prod_{\vec{r}\neq\vec{C}}r^\pi_x(\vec{r})\left\{r^\pi_z(\vec{C})r^\pi_x(\vec{C})[r^\pi_z(\vec{C})]^{-1}\right\}\nonumber\\
&=&(-1)^{2s}R^\pi_x,
\end{eqnarray}
where we have used the fact that $T_1R^\pi_xT_1^{-1}=R^\pi_x$ since $T_1$ just effectively reorders the product in the definition~(\ref{rx}) of $R^\pi_x$, which does nothing since $r^\pi_x(\vec{r})$ at different sites commute. 
We have also made use of the commutator
\begin{eqnarray}
r^\pi_z(\vec{C})r^\pi_x(\vec{C})[r^\pi_z(\vec{C})]^{-1}=(-1)^{2s}r^\pi_x(\vec{C}), 
\end{eqnarray}
which is the same as Eq.~(8) in the main text.

For the sake of completeness of the discussion, 
one could have taken the spin at the ``corner'' above to be a boundary spin as well.
The result is still the same although the derivation would have been modified a little bit in that the horizontal red bonds would be both effectively to be one lattice constant to the left of the horizontal red ones in the FIG.~\ref{gauge} and so would be the site $\vec{C}$, in addition to the unimportant adjustments of several vertical red bonds.

\section{Proofs of the symmetry of the QTM: $[\tilde{U}_g,\tilde{H}]=0$}

Here we demonstrate the symmetry of the QTM $[\tilde{U}_g,\tilde{H}]=0$ in the two different constructions, conventional and MPO-based, separately.

\subsection{Conventional construction of the QTM}

Let us consider a symmetry-twisted lattice Hamiltonian at the left top of FIG.~\ref{QTM_gauge}, where the ``$\cdots$''s on its two sides beyond the (red) twisted bond can be any complicated local interaction.
Its partition function can be represented by the QTM network below it, in which the original spatial direction is the ``time'' of QTM evolution and the original imaginary time becomes the ``real space'' of QTM Hilbert space.
Here ``$\cdots$''s in the original lattice Hamiltonian become the unknown black boxes on two sides.
For simplicity, let us define the wave function $|\tilde{\Psi}\rangle$ of the QTM Hilbert space on the domain wall (line) shown in FIG.~\ref{QTM_gauge}.
The blue boxes represent the QTM evolution $e^{-\tilde{H}}$.
Since the red bond does not affect the physics beyond it (like the black boxes),
it should be represented by an operator $\tilde{U}_g$ acted on the ``time'' slice represented by the vertical dash line in FIG.~\ref{QTM_gauge}.

\begin{figure}[h]
\centering
\includegraphics[width=15cm,pagebox=cropbox,clip]{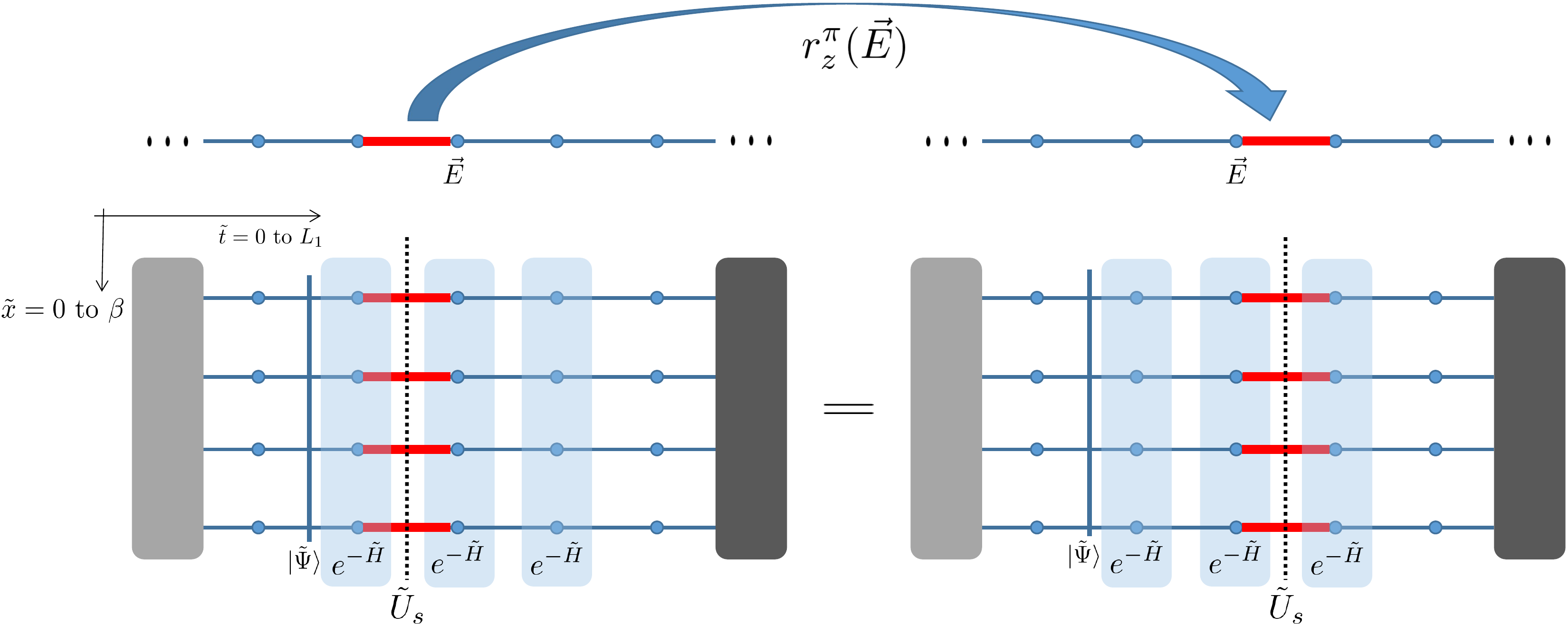}
\caption{The QTM picture of the gauge transformation $r_z^\pi(\vec{E})$ on the original Hamiltonian: the fact that the gauge transformation does not change the partition function is reflected by the commutation $[\tilde{U}_g,\tilde{H}]=0$ in the QTM paradigm.}
\label{QTM_gauge}
\end{figure}

Then, let us apply a gauge transformation $r^\pi_z(\vec{E})$ onto the original lattice Hamiltonian ${H}$, i.e., $[r^\pi_z(\vec{E})]^{-1}{H}\,r^\pi_z(\vec{E})$, so the red bond is moved by one lattice length to the right.
Such a unitary transformation is local so it does not change the ``$\cdots$'' parts far away from it, and its unitarity implies that the partition function under the transformed Hamiltonian is the same as before (that's why it is called a ``gauge'' transformation).

Translated to the QTM picture,
the ``time'' slice of $\tilde{U}_g$ is moved correspondingly and the black boxes are unchanged. 
Since the black boxes can be arbitrary, 
we arrive at
\begin{eqnarray}
e^{-\tilde{H}}e^{-\tilde{H}}\tilde{U}_ge^{-\tilde{H}}&=&e^{-\tilde{H}}\tilde{U}_ge^{-\tilde{H}}e^{-\tilde{H}}\nonumber\\
e^{-\tilde{H}}\tilde{U}_g&=&\tilde{U}_ge^{-\tilde{H}}.
\end{eqnarray} 
Furthermore,
the QTM Hamiltonian is formally defined as
\begin{eqnarray}
\tilde{H}=\ln[e^{-\tilde{H}}]=\ln\{1+[e^{-\tilde{H}}-1]\}=\text{Polynomial of }e^{-\tilde{H}}, 
\end{eqnarray}
thereby
\begin{eqnarray}
\tilde{H}\tilde{U}_g=\tilde{U}_g\tilde{H},
\end{eqnarray}
or the commutation $[\tilde{U}_g,\tilde{H}]=0$ reached.

\subsection{MPO construction of QTM}

Here we introduce a construction of QTM in terms of MPO, and demonstrate $[\tilde{U}_g,\tilde{H}]=0$ within the MPO formulation.

\begin{figure}[h]
  \centering
  \includegraphics[width=12cm,pagebox=cropbox,clip]{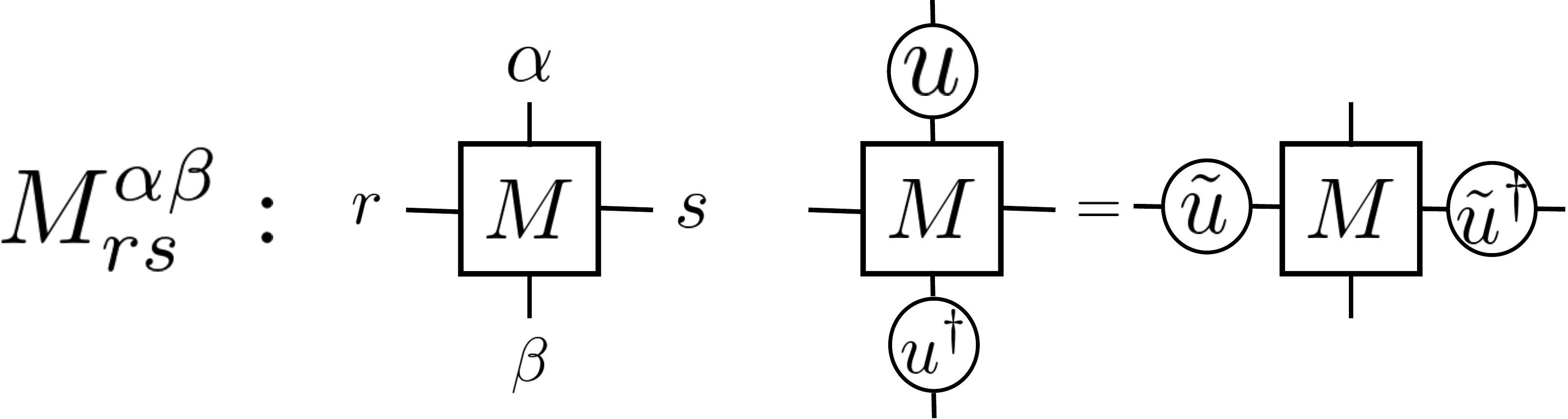}
  \caption{The tensor which is an element of the MPO. The vertical lines act on the local physical Hilbert space, while the horizontal lines do on the virtual space.}
  \label{fig:QTM-MPO-1}
\end{figure}

For illustration, here we discuss the system in one spatial dimension, although the generalization to higher dimensions is straightforward.
The ``infinitesimal'' imaginary time evolution $e^{-\delta\cdot\calH}$ can be represented by a MPO.
Let the tensor at each ``node'' as represented by a tensor $M^{\alpha \beta}_{rs}$ with the ``physical'' indices $\alpha, \beta$ and
the ``virtual'' indices $r,s$, as in Fig.~\ref{fig:QTM-MPO-1}.
The fundamental theorem on Matrix-Product Vectors (which includes the MPO by regarding the operator as an element of a vector space)
[J.~I. Cirac \textit{et. al.}, Ann. Phys. \textbf{378}, 100 (2017)]
and the  symmetry of the Hamiltonian (and thus of the MPO) imply that, for each element $g$ of the on-site symmetry group $G$, 
which transforms the local state by $u(g)$, there is a unitary
$\tilde{u}(g)$ acting on the virtual space satisfying
\begin{equation}
  {u(g)^\dagger}_{\alpha \alpha'} M^{\alpha'\beta'}_{rs} u(g)_{\beta' \beta} = \tilde{u}^\dagger(g)_{rr'} M^{\alpha\beta}_{r's'} \tilde{u}(g)_{s's} ,
\label{eq.MPO-sym}
\end{equation}
where the repeated indices are contracted, which is diagrammatically illustrated in Fig.~\ref{fig:QTM-MPO-1}.
The invariance of the MPO under $G$ immediately follows from this, as the transformations on the virtual bonds cancel upon contraction.
On the other hand, QTM in the MPO formalism is given by the cyclic ``vertical'' product of the tensor $M$ in which all the physical indices are contracted,
so that the MPO acts on the virtual indices, as in Fig.~\ref{fig:QTM-MPO-2}.

\begin{figure}[h]
  \centering
  \includegraphics[width=12cm,pagebox=cropbox,clip]{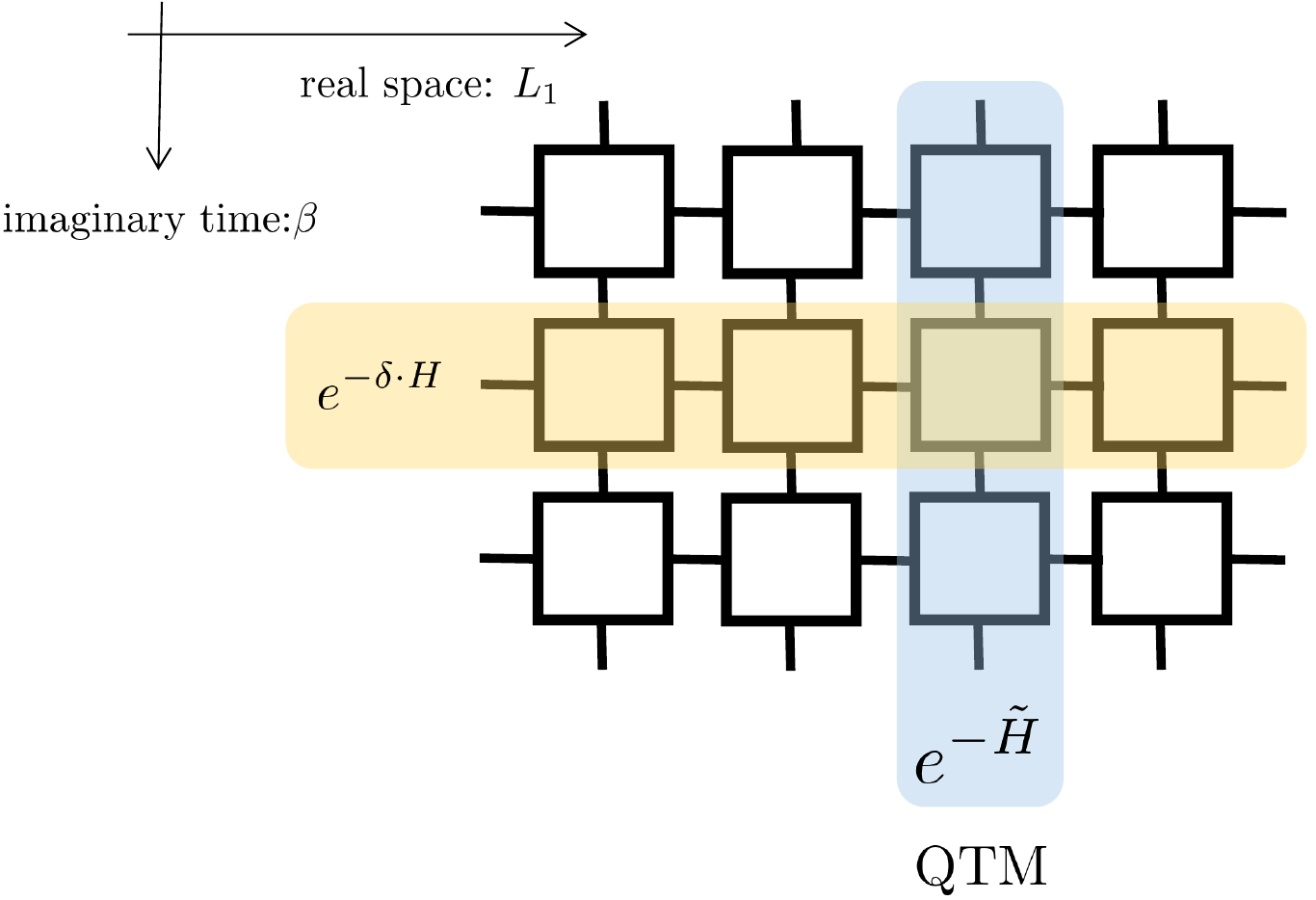}
  \caption{The MPO construction of the QTM.}
  \label{fig:QTM-MPO-2}
\end{figure}

Then the operator $\tilde{U}_g$ as defined in the text is nothing but the tensor product of $\tilde{u}(g)$ along the vertical direction.
The desired commutation relation $[\tilde{U}_g,\tilde{H}]=0$ in the main text follows directly from Eq.~\eqref{eq.MPO-sym}, as in Fig.~\ref{fig:QTM-MPO-3} below.

\begin{figure}[h]
  \centering
  \includegraphics[width=12cm,pagebox=cropbox,clip]{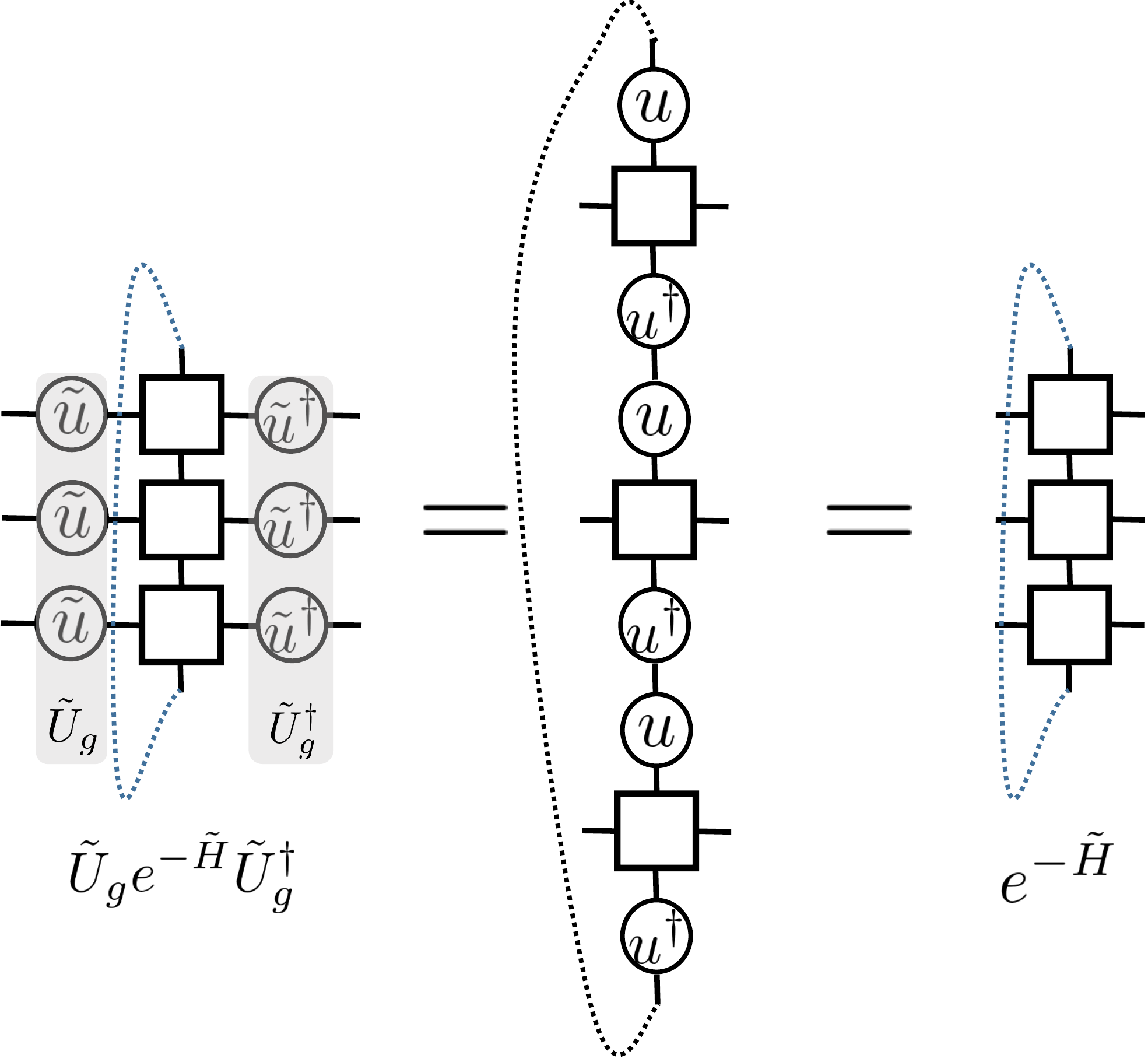}
  \caption{The diagrammatic demonstration of $[\tilde{U}_g,e^{-\tilde{H}}]=0$ or equivalently, $[\tilde{U}_g,\tilde{H}]=0$. Here the left equality uses Eq.~(\ref{eq.MPO-sym}).}
  \label{fig:QTM-MPO-3}
\end{figure}
\end{widetext}

\end{document}